\documentclass[superscriptaddress,showkeys,showpacs]{revtex4}
\usepackage{feynmf}
\usepackage{graphicx}

\begin{document}
\begin{fmffile}{fmfp}
\title{Order-$\alpha_s^2$ corrections to one-particle inclusive 
processes in DIS.}
\thanks{Partially supported by CONICET, Fundaci\'on Antorchas and ANPCyT, 
Argentina.}

\author{A. Daleo}
\email{daleo@fisica.unlp.edu.ar}
\affiliation{Laboratorio de F\'{\i}sica Te\'{o}rica\\ Departamento de
F\'{\i}sica, Facultad de Ciencias Exactas\\
Universidad Nacional de La Plata\\ C.C. 67 - 1900 La
Plata,  Argentina}
\author{C.A. Garc\'{\i}a Canal}
\affiliation{Laboratorio de F\'{\i}sica Te\'{o}rica\\ Departamento de
F\'{\i}sica, Universidad Nacional de La Plata\\ C.C. 67 - 1900 La
Plata,  Argentina}
\author{R. Sassot}
\affiliation{Departamento de F\'{\i}sica,
Universidad de Buenos Aires\\ Ciudad Universitaria, Pab.1 (1428)
Buenos Aires, Argentina}

\begin{abstract}
We analyze the order-$\alpha_s^2$ QCD corrections to semi-inclusive deep 
inelastic scattering and present results for processes initiated by a gluon.
We focus in the most singular pieces of these corrections in 
order to obtain the hitherto unknown NLO evolution kernels relevant for the 
non homogeneous QCD scale dependence of these cross sections, and to check 
explicitly factorization at this order. In so doing we discuss the 
prescription of overlapping singularities in more than one variable.
\end{abstract}

\pacs{12.38.Bx, 13.85.Ni}
\keywords{Semi-Inclusive DIS; perturbative QCD; Fracture functions}

\maketitle

\section*{Introduction}

In recent years there has been an increasing wealth of interest in 
semi-inclusive deep inelastic scattering, driven both by crucial breakthroughs 
in the QCD description of these processes\cite{ven,grau,massi,npb} and also by
an incipient availability of data encompassing polarized, unpolarized, 
leading baryon, and diffractive deep inelastic phenomena\cite{ppp}. 

From the perturbative QCD standpoint, semi-inclusive deep inelastic scattering
(SIDIS) brings before theorists two novel and interesting features. On the 
one hand,  {\em fracture functions} which, in addition to structure and 
fragmentation functions, are required for the correct description of hadrons 
produced in the forward direction  and for the factorization of collinear 
singularities. On the other, {\em non-homogeneous} Altarelli Parisi evolution 
equations, which highlights the interplay of the three intervening parton 
densities in the scale dependence of these processes\cite{ven}.  

Although the main features related to SIDIS, and specifically to fracture 
functions, have been studied at the leading order (LO) in QCD 
(order-$\alpha_s$ in the cross-section)\cite{grau,npb}, up to now no 
computations had been
done up to next to leading order (NLO) accuracy, as it is standard in the inclusive case. In particular, there were neither explicit checks of factorization
at order-$\alpha_s^2$ nor indications of how relevant the non homogeneous 
evolution might be at NLO.  

In LO, non standard evolution effects although non negligible, are restricted 
to a relatively small kinematic region, associated to the fragmentation 
configurations allowed at that order\cite{prd98}. This suggests to neglect 
these effects 
in many phenomenological analyses of polarized  SIDIS\cite{dss,ds}, 
leading baryon production \cite{prd97} and diffractive DIS\cite{prd98}, 
provided some cuts on data are introduced. 
In NLO the above mentioned kinematical restrictions are no longer present, 
which in principle may lead to important corrections. In any case, their
phenomenological relevance needs to be assessed.

From a theoretical point of view, the computation of the SIDIS NLO corrections,
and specifically the explicit check of factorization of collinear singularities
involve also some subtleties which need close attention. At variance with the 
totally inclusive case\cite{zijli}, where after a convenient integration over final states
the remaining singularities may be written as distributions in only one 
variable times a regular function, in the one particle inclusive case at 
order $\alpha_s^2$, it is necessary to keep additional variables unintegrated.
Consequently one,  must deal with entangled singularities in more than one 
variable,
corresponding for example to three particles becoming collinear simultaneously.
As usual for semi-inclusive processes, in order to check 
factorization on has to keep track of the kinematical origin or configuration 
which gives rise to
the singularity, which represents a non trivial additional complication and 
requires a detailed analysis of the singularity structure characteristic of 
the process.

In this paper we address the above mentioned issues restricting ourselves to 
processes 
where the initial state parton is a gluon. This allows to analyze and answer 
the main issues involved skipping for the moment, and for the sake of clarity,
the formidable singularity structure associated with virtual corrections to 
quark initiated processes, which will be addressed in a forthcoming 
publication.    
In doing this, we develop suitable prescription rules for dealing with
the SIDIS singularity structure.

As result of our approach we obtain the 
hitherto unknown NLO non homogeneous kernels for fracture functions and 
discuss their distinctive features 
such as their non-factorizable dependence upon two variables.
We also verify explicitly the
factorization of collinear singularities up to order $\alpha_s^2$, and give
the  expression for the renormalized fracture function in terms of the bare 
one. In order to asses the relevance of NLO corrections we compare the effects
of the new evolution kernels with the already known LO corrections. 

The outline of the paper is the following: in the next section
we introduce the relevant kinematics and conventions used and we extend the 
${\cal O}(\alpha_s)$ results for the SIDIS cross sections as required for the
later factorization of collinear singularities at ${\cal O}(\alpha_s^2)$.  
In the second section we discuss the computation of
amplitudes and phase space integration of the ${\cal O}(\alpha_s^2)$ 
processes. There, we introduce a suitable parameterization for the phase 
space of the three final state particles, and extend some of the results 
given in refs. \cite{solo,been,ingo} for the angular integration of the
corresponding amplitudes. In the third section we analyze the SIDIS 
singularity structure at order $\alpha_s^2$ and give details about the 
prescription recipes required for dealing with it. In the fourth we address 
the issue of factorization and 
technicalities associated with the convolution of distributions in many 
variables, present the novel NLO kernels and discuss the evolution of
fracture functions. In the last section we present our conclusions.
 

\section{Kinematics and ${\cal O}(\alpha_s)$ results}
We begin considering the one-particle inclusive process in which a lepton 
of momentum $l$ scatters off a nucleon of momentum $P$,
\begin{equation}
l(l)+P(P)\longrightarrow l^{\prime}(l^\prime)+h(P_h)+X,
\end{equation}
and where in addition to the emerging lepton of momentum $l^{\prime}$, a 
hadron $h$ of momentum $P_h$ is tagged in the final state. $X$ stands for 
all the unobserved particles. For simplicity we consider only the exchange 
of one photon of momentum $q=l^{\prime}-l$. In order to characterize the 
hadronic final state, in addition to the usual DIS variables
\begin{equation}
Q^2=-q^2=-(l^{\prime}-l)^2\,,\;\,\,\,\,\,\,x_{B}=\frac{Q^2}{2 P\cdot q}\,,
\;\,\,\,\,\,\,
y=\frac{P\cdot q}{P\cdot l}\,,\;\,\,\,\,\,\,S_H=(P+l)^2\,,
\end{equation}
we introduce energy and angular variables
\begin{equation}
v_{h}=\frac{E_{h}}{E_{0}(1-x_{B})}\,,\:\:\,\,\,\,\,\,\,\,\,\,\,\,\,\,
	w_{h}=\frac{1-\cos \theta_{h}}{2}\,,
\end{equation} 
where $E_{h}$ and $E_{0}$ are the energies of the produced hadron and of the
incoming proton in the $\vec{P}+\vec{q}=0$ frame, respectively. $\theta_{h}$ 
is the angle between the momenta of the hadron and the 
virtual photon in the same frame.

The corresponding cross section, differential in the final state lepton and
hadron variables, can be written as\cite{grau}
\begin{eqnarray}\label{eq:hadcsec}
\frac{d\sigma}{dx_B\, dy\, dv_h\, dw_h}=&&\sum_{i,j=q,\bar{q},g}\int_{x_B}^{1}
	\frac{du}{u}\int_{v_h}^{1} \frac{dv_j}{v_j}\int_{0}^{1}dw\,
	f_{i/P}\bigg(\frac{x_B}{u}\bigg)\,D_{h/j}\bigg(\frac{v_h}{v_j}\bigg)
	\,\frac{d\hat{\sigma}_{ij}}{dx_B\, dy\, dv_j\, dw_j}\,
	\delta(w_h-w_j)\nonumber\\
	&&+\sum_{i}\int_{\frac{x_B}{1-(1-x_B)v_h}}^{1}
	\frac{du}{u}\,\,M_{i,h/P}\left(\frac{x_B}{u},(1-x_B)v_h\right)
	\,(1-x_B)
	\,\frac{d\hat{\sigma}_{i}}{dx_B\, dy}\,\delta(1-w_h)
	\,,
\end{eqnarray}
where the sum is over all parton species. 

\begin{figure}[hbt]
\begin{minipage}{70mm}
\includegraphics[height=5cm,width=6cm]{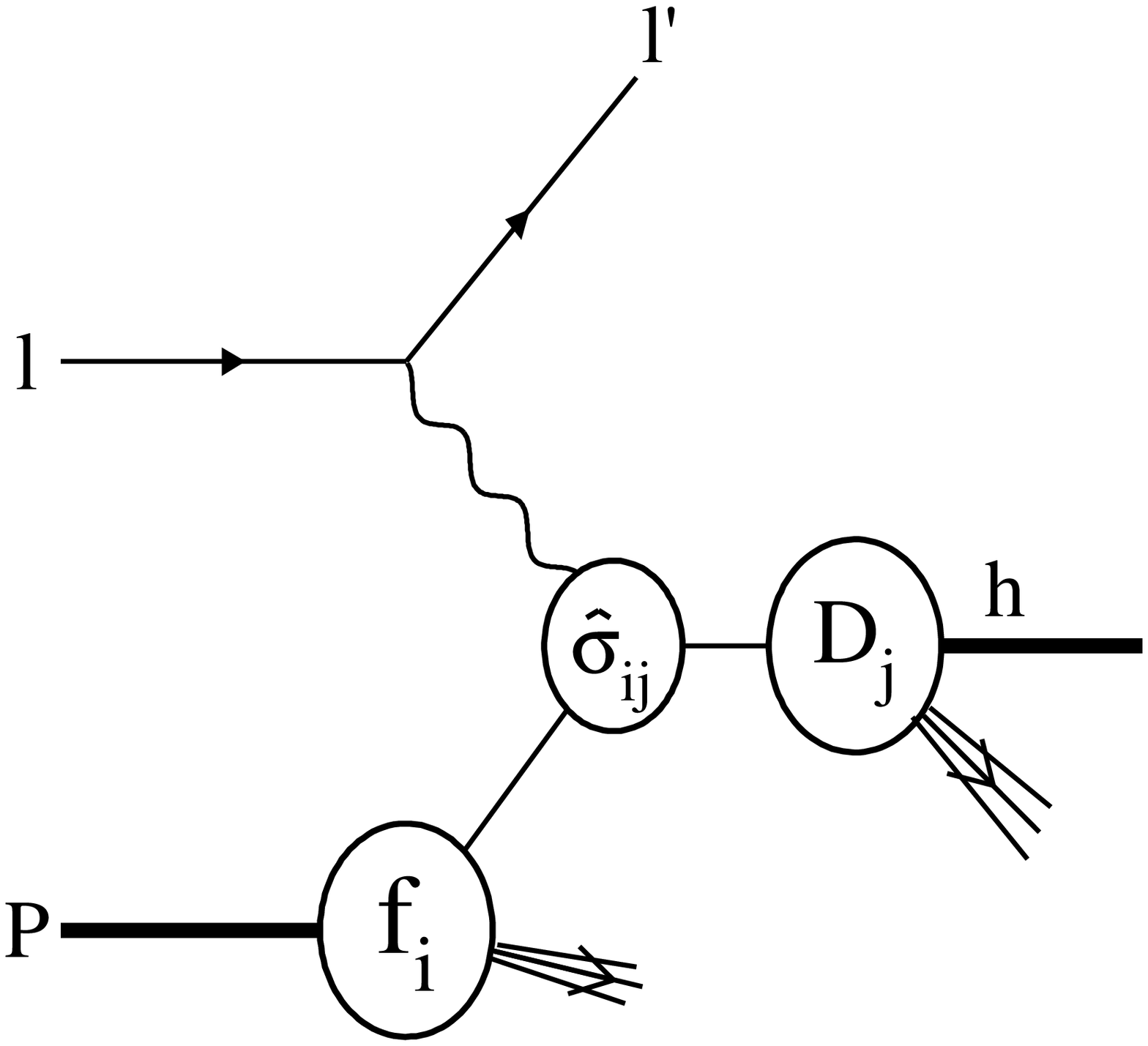}
\end{minipage}
\begin{minipage}{70mm}
\includegraphics[height=5cm,width=6cm]{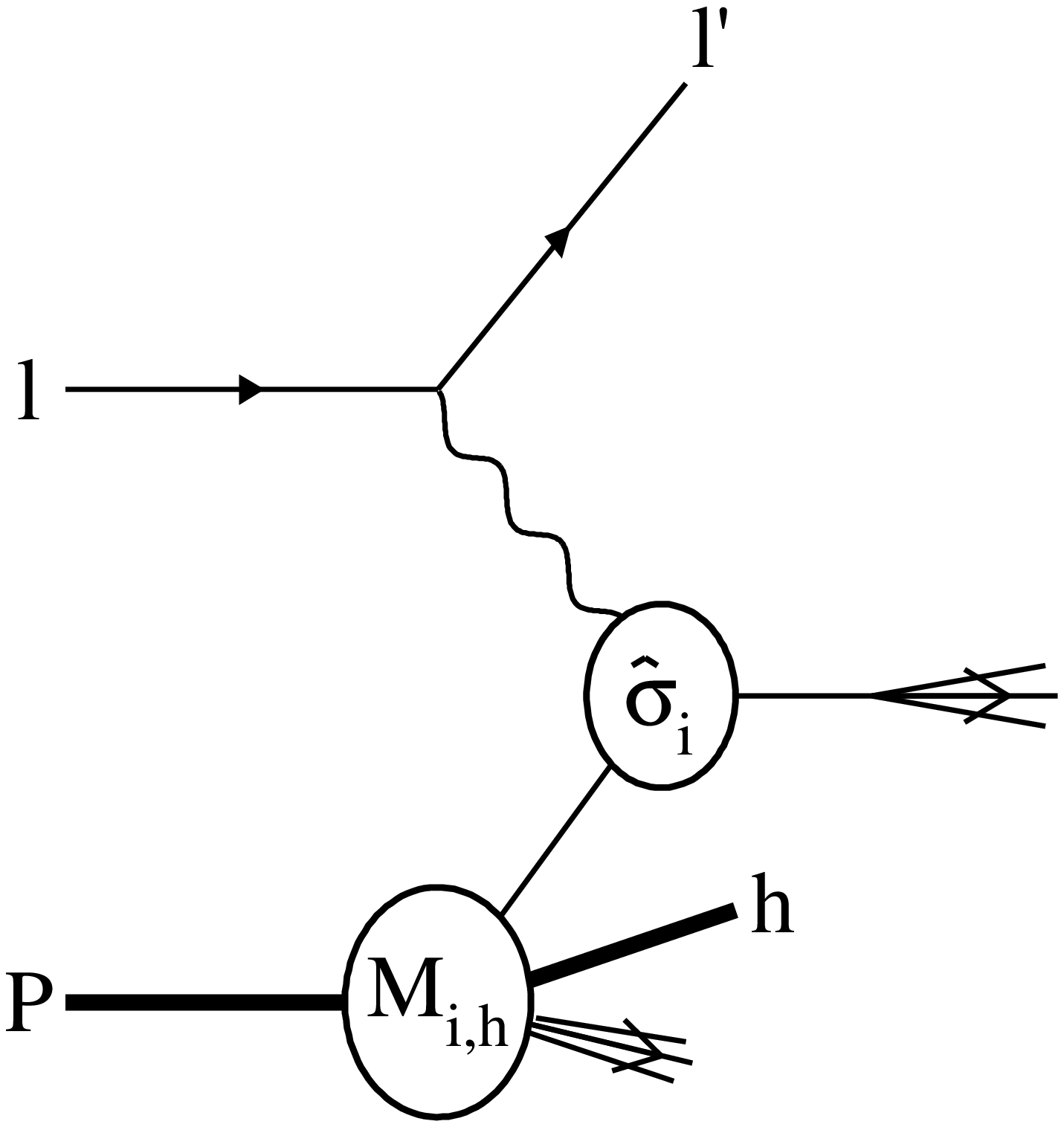}
\end{minipage}
\caption{(a) Current and (b) target fragmentation processes.}
\label{fig:frag}
\end{figure}

In the first term of the r.h.s of 
eq.(\ref{eq:hadcsec}),  $d\hat{\sigma}_{ij}$ represents the partonic 
cross section for the process $l+i\rightarrow l^{\prime}+j+X$, whereas
$f_{i/P}$ and $D_{h/j}$  are the usual partonic densities
and fragmentation functions.  The variable $u$ is related to the fraction 
of momentum of the incoming parton $\xi$ by $\xi=x_B/u$, while $v_{j}$ and 
$w_{j}$ are the partonic analogs of $v_{h}$ and $w_{h}$. 
This term, represented in fig.\ref{fig:frag}a, describes a `current fragmentation' process in which a final state parton 
$j$ fragments into the final state hadron $h$, which is produced in the 
same direction than  $j$.

In the second term of the r.h.s of eq.(\ref{eq:hadcsec}), $d\hat{\sigma}_{i}$ 
stands for the inclusive partonic cross section initiated by parton $i$ and 
is convoluted with the fracture functions  $M_{i,h/P}$. This term, shown in 
fig.\ref{fig:frag}b, corresponds to a `target 
fragmentation' process, where the initial state nucleon fragments into the
final state hadron and a parton, $i$, which participates in the hard 
scattering. In the last case, the hadron is produced in the direction of the 
incoming nucleon.

\setlength{\unitlength}{1.mm} 

The above mentioned partonic cross sections can be calculated order by 
order in perturbation theory and are related to the parton-photon 
squared matrix elements $\overline{H}^{(n)}_{\mu\nu}(i,j)$ and $\overline{H}^{(n)}_{\mu\nu}(i)$ for the $i+\gamma\rightarrow j+X$
and $i+\gamma\rightarrow X$ processes, respectively:
\begin{eqnarray}\label{eq:partxsec}
\frac{d\hat{\sigma}_{ij}}{dx_B\, dy}&=&\frac{\alpha_{em}^2}{x_B S_H}
	\left(Y_M (-g^{\mu\nu})+Y_L \frac{4x_B^2}{Q^2}P^{\mu}P^{\nu}
	\right)\frac{1}{e^2}\,\sum_{n}\overline{H}^{(n)}_{\mu\nu}(i,j) 
	J^{(n)}dv_j dw_j \nonumber\\
\frac{d\hat{\sigma}_{i}}{dx_B\, dy}&=&\frac{\alpha_{em}^2}{x_B S_H}
	\left(Y_M (-g^{\mu\nu})+Y_L \frac{4x_B^2}{Q^2}P^{\mu}P^{\nu}
	\right)\frac{1}{e^2}\,\sum_{n}\overline{H}^{(n)}_{\mu\nu}(i)
\end{eqnarray}
where  $n$ runs over the number of particles in the final state. 
Matrix elements are averaged over initial state polarizations, summed
over final state polarizations and integrated over the phase space of the 
unobserved particles. $J^{(n)}$ is the Jacobian coming from the phase space
integration and depends upon the number of final state particles $n$.
$\alpha_{em}$ stands for the fine structure constant and 
$e$ is the electron charge. Finally, $Y_M$ and $Y_L$ are the standard 
kinematic factors for the contributions of each photon polarization and 
are given by,
\begin{equation}
Y_M=\frac{1+(1-y)^2}{2y^2}\,,\:\:\:Y_L=\frac{1+4(1-y)+(1-y)^2}{2y^2}\,.
\end{equation}

\setlength{\unitlength}{1.mm} 
\begin{figure}[hbt]
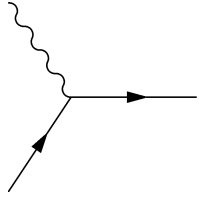

\begin{center}
\begin{minipage}{30mm}
\begin{fmfchar*}(25,25)
  \fmfstraight
  \fmfleft{qi,gamma} 
  \fmfright{qf} 
  \fmf{photon,width=5}{gamma,Pi}
  \fmfset{arrow_len}{3mm}	
  \fmf{fermion,width=5}{qi,Pi,qf}
\end{fmfchar*}
\end{minipage}
\end{center}
\caption{Born contribution to the cross sections}
\label{fig:a0}
\end{figure}

The total inclusive cross sections are well known up to order $\alpha_s^2$ 
\cite{zijli}, and more recently there have been impressive efforts to go beyond
 the NLO\cite{glov}.  
The corresponding complete expressions for the singular and finite pieces up 
to order $\alpha_s^2$ can be found in ref. \cite{zijli}. 

For the one-particle inclusive 
cross section, the zeroth-order in $\alpha_s$ comes from the 
diagram in fig.\ref{fig:a0} giving, in $d=4+\epsilon$ dimensions,  the $qq$ 
cross section:
\begin{equation}
\frac{d\hat{\sigma}^{(0)}_{qq,M}}{dx_B\,dy\,dv\,dw}=c_q\,
	\delta(1-u)\delta(1-v)\delta(w)\,, 
\end{equation}
with
\begin{equation}
c_q=\frac{\alpha^2}{2\,x_B\,S_H} 4\pi\,(2+\epsilon)\,e_q^2\,Y_M\,.
\end{equation}
The antiquark cross section $d\hat{\sigma}_{\bar{q}\bar{q}}$ is identical 
to $d\hat{\sigma}_{qq}$
whereas all the remaining cross sections vanish. The index $M$ refers to 
the metric terms in eq.(\ref{eq:partxsec}), 
longitudinal contributions 
are absent at tree level. Notice that at this order the quark is always 
produced in the backward ($w=0$) direction implying that forward hadrons 
($w=1$) would come solely from target fragmentation processes, that is, those
taken into account by fracture functions.

\setlength{\unitlength}{1.mm} 
\begin{figure}[hbt]
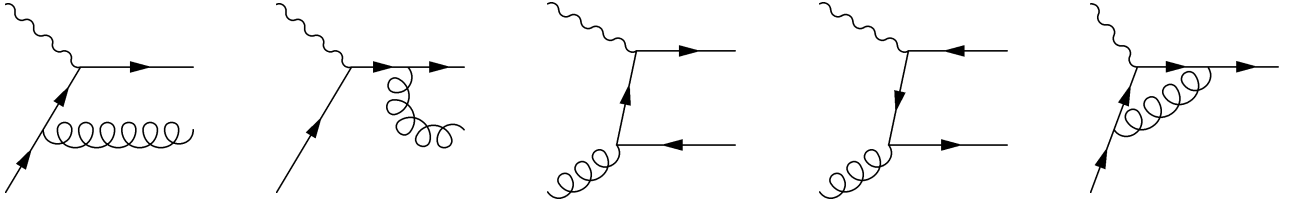

\begin{center}
\begin{minipage}{35mm}
\begin{fmfchar*}(25,25)
  \fmfstraight
  \fmfset{arrow_len}{3mm}
 \fmfleft{qi,i1,i2,gamma} 
  \fmfright{o1,g1,qf,o2} 
  \fmf{fermion,width=5}{qi,Pg,Pi,qf}
  \fmf{photon,width=5}{gamma,Pi}
  \fmffreeze
  \fmf{gluon,tension=0,width=5}{Pg,g1}
\end{fmfchar*}
\end{minipage}
\begin{minipage}{35mm}
\begin{fmfchar*}(25,25)
  \fmfstraight
  \fmfset{arrow_len}{3mm}
 \fmfleft{qi,i1,i2,gamma} 
  \fmfright{o1,g1,qf,o2} 
  \fmf{fermion,tension=0.5,width=5}{qi,Pi}
  \fmf{fermion,tension=2,width=5}{Pi,Pg,qf}
  \fmf{photon,width=5}{gamma,Pi}
  \fmffreeze
  \fmf{gluon,left,width=5}{g1,Pg}
\end{fmfchar*}
\end{minipage}
\begin{minipage}{35mm}
\begin{fmfchar*}(25,25)
  \fmfstraight  \fmfset{arrow_len}{3mm}
  \fmfleft{gi,i1,i2,i3,gamma} 
  \fmfright{o1,aqf,gf,qf,o2} 
  \fmf{photon,width=5}{gamma,vq}
  \fmf{gluon,width=5}{gi,vg1}
  \fmf{fermion,tension=0.5,width=5}{aqf,vg1}
   \fmf{plain,width=5}{vg1,vaux,vq}
  \fmf{phantom_arrow,tension=0}{vg1,vq}
  \fmf{fermion,width=5}{vq,qf}
  \fmffreeze
\end{fmfchar*}
\end{minipage}
\begin{minipage}{35mm}
\begin{fmfchar*}(25,25)
  \fmfstraight  \fmfset{arrow_len}{3mm}
  \fmfleft{gi,i1,i2,i3,gamma} 
  \fmfright{o1,aqf,gf,qf,o2} 
  \fmf{photon,width=5}{gamma,vq}
  \fmf{gluon,width=5}{gi,vg1}
  \fmf{fermion,tension=0.5,width=5}{vg1,aqf}
   \fmf{plain,width=5}{vg1,vaux,vq}
  \fmf{phantom_arrow,tension=0}{vq,vg1}
  \fmf{fermion,width=5}{qf,vq}
  \fmffreeze
\end{fmfchar*}
\end{minipage}
\begin{minipage}{35mm}
\begin{fmfchar*}(25,25)
  \fmfstraight   \fmfset{arrow_len}{3mm}
  \fmfleft{qi,gamma} 
  \fmfright{q0,q2,qf,q1} 
  \fmf{photon,tension=1,width=5}{gamma,Pi}
  \fmf{fermion,tension=1,width=5}{qi,g1,Pi}
  \fmf{fermion,tension=1,width=5}{Pi,g2,qf}
   \fmffreeze
   \fmf{gluon,width=5}{g1,g2}
\end{fmfchar*}
\end{minipage}
\end{center}
\caption{Real and virtual contributions to the $\alpha_s$ cross sections}
\label{fig:a1}
\end{figure}

The first order corrections to the one-particle inclusive cross section are 
also known. Expressions for the singular and finite terms in 
dimensional regularization\cite{Bol} can be found in \cite{grau} 
for the unpolarized case
and in \cite{npb} for the polarized one. 
In order to accomplish the factorization of collinear singularities 
at ${\cal O}(\alpha_s^2)$, one also needs the ${\cal O}(\alpha_s)$ cross 
sections up to order $\epsilon$, for this reason 
we accordingly extend here the results of \cite{grau}.
The corresponding diagrams are shown in fig.\ref{fig:a1}. 
As it is explained in ref.\cite{grau}, the integration region for the 
cross section, eq.(\ref{eq:hadcsec}), need to be splitted into two regions, B1 and B2 
respectively, in order to account for kinematical constraints in the phase 
space: 
\begin{eqnarray}\label{eq:regLO}
&&\mbox{B1}=\{u\in[x_B,x_u],\:v\in[a,1],\:w\in[0,w_r]\}\nonumber\\
&&\mbox{B2}=\{u\in[x_u,1],\:v\in[v_h,1],\:w\in[0,w_r]\}\,,
\end{eqnarray} 
with $x_u=x_B/(x_B+(1-x_B)v_h)$.
The metric terms of the unpolarized cross sections 
can be expressed in the following form:
\begin{eqnarray}
\left.d\hat{\sigma}^{(1)}_{qq(\bar{q}\bar{q}),M}\right|_{B1}
	&=&c_q\,C_{\epsilon}\left\{
	\frac{2}{\epsilon}\,P_{q\leftarrow q}^{(0)}(u)\,\delta(1-v)\,\delta(w)
	+C^{(1)}_{1qq,M}+\epsilon\,D^{(1)}_{1qq,M}
	\right\}\\
\left.d\hat{\sigma}^{(1)}_{qg(\bar{q}g),M}\right|_{B1}
	&=&c_q\,C_{\epsilon}\left\{
	\frac{2}{\epsilon}\,P_{qg\leftarrow q}^{(0)}(u)\,\delta(a-v)\,
	\delta(1-w)
	+C^{(1)}_{1qg,M}+\epsilon\,D^{(1)}_{1qg,M}
	\right\}\\
\left.d\hat{\sigma}^{(1)}_{gq(g\bar{q}),M}\right|_{B1}
	&=&c_q\,C_{\epsilon}\left\{
	\frac{2}{\epsilon}\,P_{\bar{q}q\leftarrow g}^{(0)}(u)\,
	\delta(v-a)\,\delta(1-w)
	+\frac{2}{\epsilon}\,P_{q\leftarrow g}^{(0)}(u)\,
	\delta(1-v)\,\delta(w)+
	C^{(1)}_{1gq,M}+\epsilon\,D^{(1)}_{1gq,M}
	\right\}\\
\left.d\hat{\sigma}^{(1)}_{qq(\bar{q}\bar{q}),M}\right|_{B2}
	&=&c_q\,C_{\epsilon}\left\{
	\frac{2}{\epsilon}\,\left(P_{q\leftarrow q}^{(0)}(u)\,\delta(1-v)
	+P_{q\leftarrow q}^{(0)}(v)\,\delta(1-u)\right)\delta(w)
	+C^{(1)}_{2qq,M}+\epsilon\,D^{(1)}_{2qq,M}
	\right\}\\
\left.d\hat{\sigma}^{(1)}_{qg(\bar{q}g),M}\right|_{B2}
	&=&c_q\,C_{\epsilon}\left\{
	\frac{2}{\epsilon}\,P_{g\leftarrow q}^{(0)}(v)\,\delta(1-u)\,\delta(w)
	+C^{(1)}_{2qg,M}+\epsilon\,D^{(1)}_{2qg,M}
	\right\}\\
\left.d\hat{\sigma}^{(1)}_{gq(g\bar{q}),M}\right|_{B2}
	&=&c_q\,C_{\epsilon}\left\{
	\frac{2}{\epsilon}\,P_{q\leftarrow g}^{(0)}(u)\,
	\delta(1-v)\,\delta(w)+
	C^{(1)}_{2gq,M}+\epsilon\,D^{(1)}_{2gq,M}
	\right\}
\end{eqnarray}
where $C_{\epsilon}$ is defined by
\begin{equation}
C_{\epsilon}=\frac{\alpha_s}{2\pi}\,f_{\Gamma}
	\left(\frac{Q^2}{4\pi\mu^2}\right)^
	{\epsilon/2},\:\:\:\:f_{\Gamma}={\small
	\frac{\Gamma(1+\epsilon/2)}{\Gamma(1+\epsilon)}}
\end{equation}
and
\begin{equation}\label{eq:aywr}
a=\frac{(1-u)\,x_B}{u\,(1-x_B)}\,,\:\:\:
	w_r=\frac{(1-v)\,(1-u)\,x_B}{v\,(u-x_B)}\,.
\end{equation}
The $P^{(0)}_{i\leftarrow j}$ are the usual LO Altarelli-Parisi kernels
\cite{AP}, 
whereas the $P^{(0)}_{jk\leftarrow i}$ are the unsubstracted ones. 
Expressions for them and for the coefficient functions $C^{(1)}$ 
can be found in Appendix B of \cite{grau}. Notice that in order to match our
notation the results of ref.\cite{grau} should be multiplied by a 
factor $\delta(w_r-w)$. The coefficients $D^{(1)}$ are explicitly given in 
Appendix I below. Similar expressions for the longitudinal 
cross sections, which are finite at this order, can be obtained from the 
results in ref. \cite{grau}.   

Notice that at this order, the fragmenting parton can be produced in any 
direction, including the forward one, but all singular contributions come either from the
backward or from the forward direction. The former terms are factorized into 
partonic densities and fragmentation functions, whereas the forward 
singularities can only be factorized in the redefinition of fracture functions.
This factorization gives rise, as we have already mentioned, 
to non-homogeneous terms in the evolution equations of fracture 
functions \cite{ven}. Notice also that 
singular terms in the forward direction are always accompanied by a 
$\delta(v-a)$ factor. This is a characteristic feature of the LO results
that in general, will not be present at the $\alpha_s^2$ order.


\section{${\cal O}(\alpha_s^2)$ amplitudes and phase space integration.}  

In this section we outline the computation of the order $\alpha_s^2$ 
amplitudes for the one-particle inclusive cross sections. At this order
the relevant amplitudes have either two or three final state partons, 
related to virtual and real contributions, respectively. We have computed the 
corresponding hadronic tensors $H_{\mu\nu}$ in $d=4+\epsilon$ 
dimensions, in the Feynman gauge, and taking all the quarks to be massless. 

Algebraic manipulations were 
performed with the aid of the program {\sc Mathematica} \cite{math} and 
the package {\sc Tracer} \cite{tracer} to perform the traces over the Dirac indices. 
In order to obtain the one-particle inclusive cross sections, one has
to integrate the resulting matrix elements over the internal loop momenta 
and over the phase space of the unobserved particles in the final state, which 
is one of the hardest and most delicate parts of the calculation. 
 
\begin{figure}[h]
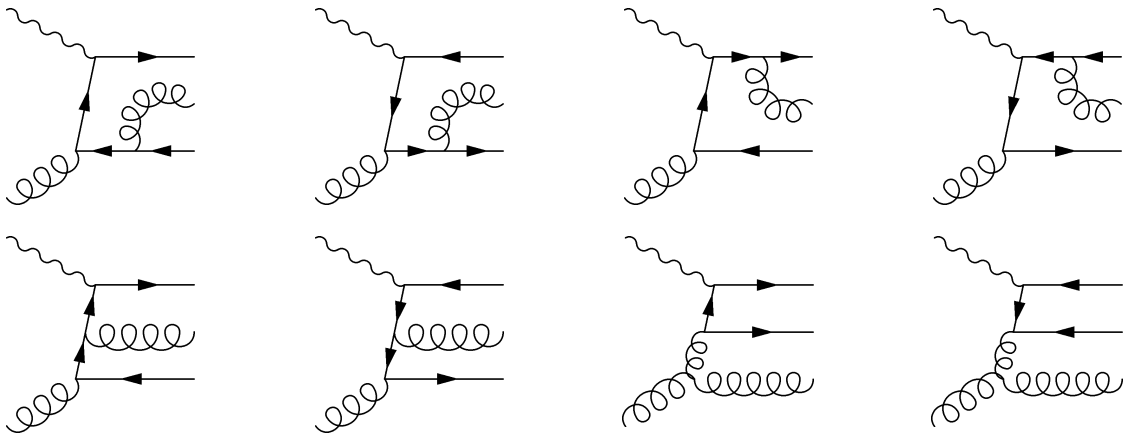

\begin{center}
\begin{minipage}{40mm}
\begin{fmfchar*}(25,25)
  \fmfstraight  \fmfset{arrow_len}{3mm}
  \fmfleft{gi,i1,i2,i3,gamma} 
  \fmfright{o1,aqf,gf,qf,o2} 
  \fmf{photon,width=5}{gamma,vq}
  \fmf{gluon,width=5}{gi,vg1}
  \fmf{fermion,width=5}{aqf,vg2,vg1}
   \fmf{plain,width=5}{vg1,vaux,vq}
  \fmf{phantom_arrow,tension=0}{vg1,vq}
  \fmf{fermion,width=5}{vq,qf}
  \fmffreeze
  \fmf{gluon,left,width=5}{vg2,gf}
\end{fmfchar*}
\end{minipage}
\begin{minipage}{40mm}
\begin{fmfchar*}(25,25)
  \fmfstraight   \fmfset{arrow_len}{3mm}
  \fmfleft{gi,i1,i2,i3,gamma} 
  \fmfright{o1,aqf,gf,qf,o2} 
  \fmf{photon,width=5}{gamma,vq}
  \fmf{gluon,width=5}{gi,vg1}
  \fmf{fermion,width=5}{vg1,vg2,aqf}
   \fmf{plain,width=5}{vg1,vaux,vq}
  \fmf{phantom_arrow,tension=0}{vq,vg1}
  \fmf{fermion,width=5}{qf,vq}
  \fmffreeze
  \fmf{gluon,left,width=5}{vg2,gf}
\end{fmfchar*}
\end{minipage}
\begin{minipage}{40mm}
\begin{fmfchar*}(25,25)
  \fmfstraight  \fmfset{arrow_len}{3mm}
  \fmfleft{gi,i1,i2,i3,gamma} 
  \fmfright{o1,aqf,gf,qf,o2} 
  \fmf{photon,width=5}{gamma,vq}
  \fmf{gluon,width=5}{gi,vg1}
  \fmf{fermion,tension=0.5,width=5}{aqf,vg1}
   \fmf{plain,width=5}{vg1,vaux,vq}
  \fmf{phantom_arrow,tension=0}{vg1,vq}
  \fmf{fermion,tension=2,width=5}{vq,vg2,qf}
  \fmffreeze
  \fmf{gluon,left,width=5}{gf,vg2}
\end{fmfchar*}
\end{minipage}
\begin{minipage}{40mm}
\begin{fmfchar*}(25,25)
  \fmfstraight  \fmfset{arrow_len}{3mm}
  \fmfleft{gi,i1,i2,i3,gamma} 
  \fmfright{o1,aqf,gf,qf,o2}
  \fmf{photon,width=5}{gamma,vq}
  \fmf{gluon,width=5}{gi,vg1}
  \fmf{fermion,tension=0.5,width=5}{vg1,aqf}
   \fmf{plain,width=5}{vg1,vaux,vq}
  \fmf{phantom_arrow,tension=0}{vq,vg1}
  \fmf{fermion,tension=2,width=5}{qf,vg2,vq}  
  \fmffreeze
  \fmf{gluon,left,width=5}{gf,vg2}
\end{fmfchar*}
\end{minipage}

\vspace{0.5cm}

\begin{minipage}{40mm}
\begin{fmfchar*}(25,25)
  \fmfstraight  \fmfset{arrow_len}{3mm}
  \fmfleft{gi,i1,i2,i3,gamma} 
  \fmfright{o1,aqf,gf,qf,o2} 
  \fmf{photon,width=5}{gamma,vq}
  \fmf{gluon,width=5}{gi,vg1}
  \fmf{fermion,tension=0.5,width=5}{aqf,vg1}
   \fmf{fermion,width=5}{vg1,vg2,vq}
  \fmf{fermion,width=5}{vq,qf}
  \fmffreeze
  \fmf{gluon,width=5}{vg2,gf}
\end{fmfchar*}
\end{minipage}
\begin{minipage}{40mm}
\begin{fmfchar*}(25,25)
  \fmfstraight  \fmfset{arrow_len}{3mm}
  \fmfleft{gi,i1,i2,i3,gamma} 
  \fmfright{o1,aqf,gf,qf,o2} 
  \fmf{photon,width=5}{gamma,vq}
  \fmf{gluon,width=5}{gi,vg1}
  \fmf{fermion,tension=0.5,width=5}{vg1,aqf}
   \fmf{fermion,width=5}{vq,vg2,vg1}
  \fmf{fermion,width=5}{qf,vq}
  \fmffreeze
  \fmf{gluon,width=5}{vg2,gf}
\end{fmfchar*}
\end{minipage}
\begin{minipage}{40mm}
\begin{fmfchar*}(25,25)
  \fmfstraight  \fmfset{arrow_len}{3mm}
  \fmfleft{gi,i1,i2,i3,gamma} 
  \fmfright{o1,aqf,gf,qf,o2} 
  \fmf{photon,width=5}{gamma,vq}
  \fmf{gluon,width=5}{vg2,vg1,gi}
  \fmf{gluon,tension=0.5,width=5}{vg1,aqf}
   \fmf{fermion,width=5}{vg2,vq,qf}
  \fmffreeze
  \fmf{fermion,width=5}{vg2,gf}
\end{fmfchar*}
\end{minipage}
\begin{minipage}{40mm}
\begin{fmfchar*}(25,25)
  \fmfstraight  \fmfset{arrow_len}{3mm}
  \fmfleft{gi,i1,i2,i3,gamma} 
  \fmfright{o1,aqf,gf,qf,o2} 
  \fmf{photon,width=5}{gamma,vq}
  \fmf{gluon,width=5}{vg2,vg1,gi}
  \fmf{gluon,tension=0.5,width=5}{vg1,aqf}
   \fmf{fermion,width=5}{qf,vq,vg2}
  \fmffreeze
  \fmf{fermion,width=5}{gf,vg2}
\end{fmfchar*}
\end{minipage}
\end{center}
\caption{Real contributions to the $\alpha_s^2$ cross sections}
\label{fig:a2real}
\end{figure}

As we mentioned, in the present paper we restrict ourselves to gluon initiated processes,
the corresponding ${\cal O}(\alpha_s^2)$ real contributions ($gg$ and $gq$ 
processes) are shown in fig.\ref{fig:a2real}. 
For the first of these processes, the phase space integration is over 
the momenta of the quark-antiquark pair, whereas for the second it is over
the gluon-antiquark momenta. To perform this integrals, we choose to work 
in the center of mass frame of the two unobserved partons and get for 
the phase space:
\begin{eqnarray}
dPS^{3}&=&
	\frac{1}{Q^2}\,\frac{1}{\Gamma\left(1+\epsilon\right)}
	\left(\frac{Q^2}{4\pi}\right)^{\epsilon}\,(4\pi)^4
	\left(\frac{1-x_B}{x_B}\right)^{1+\epsilon/2}
	\left(\frac{u-x_B}{x_B}\right)^{\epsilon/2}\,
	v^{1+3\,\epsilon/2}
	\theta(w_r-w)\,
	(w_r-w)^{\epsilon/2}\, w^{\epsilon/2}\nonumber\\
	&&\times(1-w)^{\epsilon/2}
	dv\,dw\,\sin^{1+\epsilon}\beta_{1}\,\sin^{\epsilon}\beta_{2}\,
	d\beta_{1}\,d\beta_{2}\,.
\end{eqnarray}
The angles $\beta_1$ and $\beta_2$ are the polar and azimuthal angles of
one of the unobserved partons defined in the mentioned frame. The 
orientation of the axes in this frame was chosen in order to simplify 
the functions to be integrated. $v$ and $w$ are the energy and the 
angle of the fragmenting parton, respectively. In this case, 
as we anticipated, there is no correlation between them, but the 
$\theta$ function splits the integration region $R$ in the $u$, $v$ 
and $w$ volume: 
$R=\mbox{B0}\cup \mbox{B1}\cup \mbox{B2}$ 
where $\mbox{B0}$ is given by 
\begin{equation}
\mbox{B0}=\{u\in[x_B,x_u],\:v\in[v_h,a],\:w\in[0,1]\}\,.
\end{equation} 
and $\mbox{B1}$ and $\mbox{B2}$ are the LO regions given in 
eq.(\ref{eq:regLO}). 

As it is common practice, in order to perform the angular integration, 
matrix elements have to be decomposed, via partial fractioning, in such 
a way that all the angular integrals end in the standard form \cite{been} 
\begin{equation}\label{eq:iang}
I(k,l)=\int_{0}^{\pi}d\beta_1\int_{0}^{\pi}d\beta_2\,
	\frac{\sin^{1+\epsilon}\beta_{1}
	\sin^{\epsilon}\beta_{2}}
	{(a+b\cos\beta_{1})^{k}(A+B\cos\beta_{1}+
	C\sin\beta_{1}\cos\beta_{2})^{l}}\,.
\end{equation}
This kind of  integrals can be classified into four categories according to 
whether their parameters satisfy either $a^2=b^2$ or $A^2=B^2+C^2$, both 
relations simultaneously, or neither of them.
In the present case, after the partial fractioning we obtained 31 independent 
integrals. 23 of these integrals were calculated to all orders in $\epsilon$ 
extending the results of refs. \cite{solo,been,ingo}. 
The remaining 8, which we were not able to calculate
to all orders, need to be carefully handled before expanding them
in a power series in $\epsilon$.

The difficulty with the above mentioned integrals is that $\epsilon$
is not only regulating the $\beta$ integration, but also the singularities 
in the remaining variables: $u$, $v$ and $w$.
Although the integrals may be regular functions of these variables, their
coefficients may be not. An illustrative example of this situation is
given by the integral $I(1,1)|_{A^2=B^2+C^2}$, as in eq.(\ref{eq:iang})  with $k=1$, $l=1$, and satisfying $A^2=B^2+C^2$, for which the
order by order computation in $\epsilon$ gives (eq. C30 in \cite{been})
\begin{equation}\label{eq:ejint1}
\left.I(1,1)\right|_{A^2=B^2+C^2}=\frac{\pi}{a\,A-b\,B}\left\{\frac{2}
	{\epsilon}+\log\left[\frac{(a\,A-b\,B)^2}{(a^2-b^2)A^2}\right]
	+{\cal O}(\epsilon)\right\}\,.
\end{equation}
Let us first consider the case when $a+b\sim (1-w)$. As long as the 
coefficient 
of this integral is regular at $w=1$, the above expression is integrable
as function of $w$. However, if the coefficient has an extra factor of
$(1-w)^{-1}$ the resulting expression is ill defined due to the logarithm 
in eq.(\ref{eq:ejint1}) which behaves as $\log(1-w)$. 
In order to skip this problem, one can recast eq.(\ref{eq:iang}) using
the general methods described in the appendix A of ref.\cite{solo}
obtaining:
\begin{eqnarray}
\left.I(1,1)\right|_{A^2=B^2+C^2}&=&-\int_{0}^{1}dx
	\frac{\pi\,(1-x)^{-1+\epsilon/2}}{A\,(a-b)+b\,(A-B)\,x}
	\left\{\left(\frac{A}{A\,(a-b)+b\,(A-B)\,x}\right)^{\epsilon/2}
	\right.\nonumber\\
	&\times&\left.(a+b)^{\epsilon/2}
	\,_{2}F_{1}\left[\frac{\epsilon}{2},\epsilon,1+\epsilon;
	\frac{b\,(A-B)\,-2A\,b}{A\,(a-b)+b\,(A-B)\,x}\right]-2
	\right\}\,.
\end{eqnarray}
The integral in last equation can be splitted into two pieces, 
one containing the hypergeometric function, which can be integrated 
order by order in a power expansion in $\epsilon$ after factoring out 
$(a+b)^{\epsilon/2}$, and the other
which can be integrated to all orders in $\epsilon$: 
\begin{eqnarray}
\left.I(1,1)\right|_{A^2=B^2+C^2}&=&\frac{\pi}{a\,A-b\,B}\left\{\frac{4}
	{\epsilon}+2\log\left[\frac{a\,A-b\,B}{(a-b)A}\right]\right.
	+\left.
	(a+b)^{\epsilon/2}\left[-\frac{2}{\epsilon}+\log[a-b]\right]
	+{\cal O}(\epsilon)\right\}\,.
\end{eqnarray}
Notice that factoring out $(a+b)^{\epsilon/2}$ before the expansion in 
powers of $\epsilon$ avoids the appearance of powers of $\log(a+b)$ in the 
series, which would be singular in the $w\rightarrow 1$ limit. 
In this way we obtain well defined integrals in $w$, 
as long as $\epsilon> 0$, even if the coefficient has a pole in $w=1$, 
which is rather frequent.
In some cases, the angular integrals have singularities in $u$, $v$ or $w$ 
by themselves, but they can be managed in the same way as in the example 
above. 

The procedure just illustrated was performed for all the 8 integrals and for
the different combinations of singularities in $u$, $v$ and $w$, expressions 
for them are available upon request.
It is also important to stress that, as the singular distributions that show 
up in the matrix elements after the angular integration give rise to 
additional poles in $\epsilon$, it is  necessary to calculate contributions
 up to order $\epsilon^3$ in the angular integrals. Fortunately, these poles 
are always accompanied by one or more $\delta$ functions and those higher 
order terms only need to be calculated in the corresponding limits, what 
simplifies considerably the integrals.  

\begin{figure}[ht]
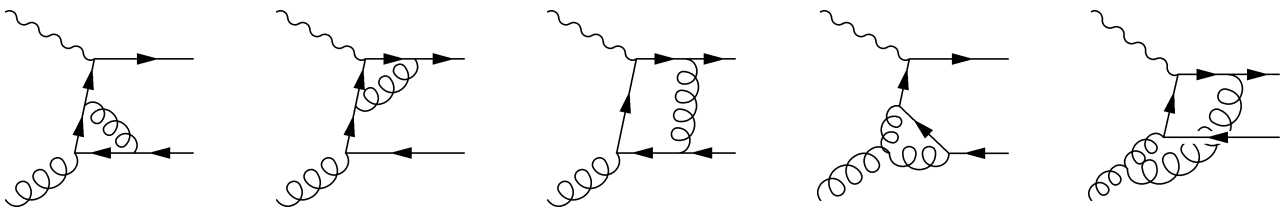

\begin{center}
\begin{minipage}{35mm}
\begin{fmfchar*}(25,25)
  \fmfstraight  \fmfset{arrow_len}{3mm}
  \fmfleft{gi,i1,i2,i3,gamma} 
  \fmfright{o1,aqf,gf,qf,o2} 
  \fmf{photon,width=5}{gamma,vq}
  \fmf{gluon,width=5}{gi,vg1}
  \fmf{fermion,width=5}{aqf,vg2,vg1}
   \fmf{fermion,width=5}{vg1,vaux,vq}
  \fmf{fermion,width=5}{vq,qf}
  \fmffreeze
  \fmf{gluon,width=5}{vg2,vaux}
  \fmf{phantom,left}{vg2,gf}
\end{fmfchar*}
\end{minipage}
\begin{minipage}{35mm}
\begin{fmfchar*}(25,25)
  \fmfstraight  \fmfset{arrow_len}{3mm}
  \fmfleft{gi,i1,i2,i3,gamma} 
  \fmfright{o1,aqf,gf,qf,o2}
  \fmf{photon,width=5}{gamma,vq}
  \fmf{gluon,width=5}{gi,vg1}
  \fmf{fermion,tension=0.5,width=5}{aqf,vg1}
   \fmf{fermion,width=5}{vg1,vaux,vq}
  \fmf{fermion,tension=2,width=5}{vq,vaux2,qf}
  \fmffreeze
  \fmf{gluon,width=5}{vaux,vaux2}
  \fmf{phantom,left}{vg2,gf}
\end{fmfchar*}
\end{minipage}
\begin{minipage}{35mm}
\begin{fmfchar*}(25,25)
  \fmfstraight  \fmfset{arrow_len}{3mm}
  \fmfleft{gi,i1,i2,i3,gamma} 
  \fmfright{o1,aqf,gf,qf,o2}
  \fmf{photon,width=5}{gamma,vq}
  \fmf{gluon,width=5}{gi,vg1}
  \fmf{fermion,width=5}{aqf,vg2,vg1}
   \fmf{plain,width=5}{vg1,vaux,vq}
  \fmf{phantom_arrow,tension=0}{vg1,vq}
  \fmf{fermion,tension=2,width=5}{vq,vaux2,qf}
  \fmffreeze
  \fmf{gluon,width=5}{vg2,vaux2}
\end{fmfchar*}
\end{minipage}
\begin{minipage}{35mm}
\begin{fmfchar*}(25,25)
  \fmfstraight  \fmfset{arrow_len}{3mm}
  \fmfleft{gi,i1,i2,i3,gamma} 
  \fmfright{o1,aqf,gf,qf,o2} 
  \fmf{photon,width=5}{gamma,vq}
  \fmf{gluon,width=5}{vg1,gi}
  \fmf{fermion,width=5}{aqf,vg2}
   \fmf{phantom}{vg2,vg1}
   \fmf{phantom}{vg1,vaux}
   \fmf{fermion,width=5}{vaux,vq}
  \fmf{fermion,width=5}{vq,qf}
  \fmffreeze
  \fmf{fermion,width=5}{vg2,vaux}
  \fmf{gluon,width=5}{vg1,vg2}
  \fmf{gluon,width=5}{vaux,vg1}
  \fmf{phantom,left}{vg2,gf}
\end{fmfchar*}
\end{minipage}
\begin{minipage}{35mm}
\begin{fmfchar*}(25,25)
  \fmfstraight  \fmfset{arrow_len}{3mm}
  \fmfleft{gi,i1,i2,gamma} 
  \fmfright{o1,aqf,gf,o4,qf,o3,o2} 
  \fmf{photon,width=5}{gamma,vq}
   \fmf{phantom,tension=2}{gi,v3g,vg1}
  \fmf{phantom,tension=0.5}{gf,vg1}
   \fmf{plain,width=5}{vg1,vq}
  \fmf{phantom_arrow,tension=0}{vg1,vq}
  \fmf{plain,tension=2,width=5}{vq,vqg,qf}
  \fmffreeze
  \fmf{fermion,width=5}{vq,vqg,qf}
 \fmf{fermion,rubout=8.,width=5}{gf,vg1}
 \fmf{gluon,width=5}{vg1,v3g,gi}
  \fmf{gluon,right=.4,width=5}{v3g,vqg}
\end{fmfchar*}
\end{minipage}
\end{center}
\caption{One loop contributions to the $\alpha_s^2$ $gq$ cross section. 
Diagrams obtained from the first four by reversing the quark line must
be also taken into account.}
\label{fig:a2virt}
\end{figure}

Virtual contributions for the $gq$ subprocess are obtained from the 
interference of the one loop graphs in fig.\ref{fig:a2virt} with the 
box graphs in fig.\ref{fig:a1}. Integration over the loop momentum 
was done using the standard Passarino-Veltman \cite{pasa} reduction algorithm
and computing the resulting 2, 3 and 4-point scalar integrals.
The integration over the phase space of the unobserved antiquark can be 
trivially performed using the energy-momentum conservation $\delta$ 
function. After this 
integration, the remaining phase space can be written as
\begin{eqnarray}\label{eq:PS2}
dPS^{(2)}&=&\frac{1}{8\pi\,\Gamma\left(1+\epsilon/2\right)}
	\left(\frac{Q^2}{4\pi}\right)^{\epsilon/2}
	\frac{u\,(1-x_B)}{u-x_B}\left(\frac{1-x_B}{x_B}\right)^{\epsilon/2}
	 v^{\epsilon}w^{\epsilon/2}(1-w)^{\epsilon/2}\,
	\delta(w_r-w)\,dv\,dw\,,
\end{eqnarray}
where $v$ and $w$ are the energy and angular variables of the
hadronizing quark respectively. As can be seen from the $\delta$ function in 
eq.(\ref{eq:PS2}), these two variables are correlated, which  is a 
distinctive feature of the two particle phase space. It implies an
additional constraint over the integration region: as $w=w_r\le 1$ then 
$v\ge a$ should be satisfied and, as it happens for the order $\alpha_s$ 
corrections, 
the integration region $V$ has to be splitted into 
$V=\mbox{B1}\cup\mbox{B2}$ where
$\mbox{B1}$ and $\mbox{B2}$ have been already defined in eq.(\ref{eq:regLO}).


\section{Singularity structure}

Once the angular integrations are performed, the hadronic tensor shows
a rich variety of singularities in the $(u,v,w)$ space, regulated by the
parameter $\epsilon$. As it is standard in this kind of calculations, the
above mentioned  singularities should be {\em prescribed} in order 
to get a series expansion in powers of $\epsilon$ suitable for making 
explicit their cancellation. These cancellations are performed  
by coupling constant renormalization for the UV singularities, by
cancellations between virtual and real contributions for the soft ones, 
and by renormalization of
parton densities, fragmentation and fracture functions in the collinear case. 

A standard example for the above mentioned {\em prescriptions} is the 
appearance of 
factors like $(1-u)^{-1+\epsilon}$ in the totally inclusive cross section 
where, after the phase space integration, $u$ is the only remaining variable. 
In this case one can use the standard substitution:
\begin{equation}\label{eq:prescu1}
(1-u)^{-1+\epsilon}\equiv \frac{1}{\epsilon}\,\delta(1-u)+
	\left(\frac{1}{1-u}\right)_{+u[0,\underline{1}]}
	+{\cal O}(\epsilon)\,,
\end{equation}  
where $(1/(1-u))_{+u[0,\underline{1}]}$ is the usual `plus' distribution:
\begin{equation}
\int_{0}^{1}du \left(\frac{1}{1-u}\right)_{+u[0,\underline{1}]}f(u)=
	\int_{0}^{1}du\,\frac{f(u)-f(1)}{1-u}\,.
\end{equation}
However, in the one-particle inclusive case, the structure of the 
singularities is much more complex, mixing the three variables and consequently
this simple prescription is no longer adequate. 
\begin{figure}[ht]
\begin{center}
\includegraphics[height=7cm,width=7cm]{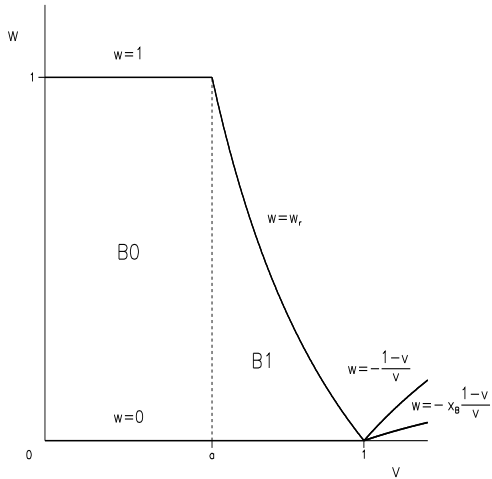}
\end{center}
\caption{Position of the singularities in the $v$-$w$ plane for 
$x_B\le u\le x_{u}$. 
The bold lines represent the curves where 
the hadronic tensor becomes singular.}\label{fig:muzub}
\end{figure}
In fig.(\ref{fig:muzub}) we show the curves along which the 
singularities in the regions $\mbox{B0}$ and 
$\mbox{B1}$ appear in the $v$-$w$ plane after the angular integration 
is performed. We will focus on this two regions because 
they contain all the 
singularities in the forward direction which need to be factorized in
the redefinition of fracture functions. 
The case of region $\mbox{B2}$ is quite similar to that of
$\mbox{B1}$ without the complications of the poles in $w=w_r=1$ but 
with additional singularities along the plane $u=1$, and it will be 
discussed at the end of this section. 

A simple inspection of fig.(\ref{fig:muzub}) allows one to distinguish 
different 
possibilities for the singularity structure of the terms in the hadronic 
tensor. In principle the integration leads to terms that can have none, one, or
two poles along the thick curves in the figure, respectively.

The case of a single pole can easily be handled with minor 
modifications to the prescription formula in eq.(\ref{eq:prescu1}). 
For terms with more than a single pole, the singular curves 
can either intersect themselves (for example poles in $w=1$ and $w=w_r$)  
or not (for instance $w=0$ and $w=1$).

Terms with poles along two non-intersecting curves can be shown to be always 
transformed by partial fractioning into two terms with single poles, which 
reduce to the previous case. On the other 
hand, the case of two intersecting singular curves can not be reduced to a 
simpler one and needs to be treated in a more subtle way.

Overlapping singularities as those mentioned  in the previous paragraph can be
further classified according to whether: a)   
both curves lie in the integration region, like
$w=0$ and $w=w_r$ in region $\mbox{B1}$; b) one of them comes from the 
outside of the integration region but intersects it at some point, as it is 
the case of $w=1$ and $w=w_r$ in $\mbox{B0}$; and c) both curves converge into a 
single point of the integration region but coming from the outside, like 
$w=-(1-v)/v$ and $w=-x_B(1-v)/v$. 

The first and third occurrences can be cast into the second, by partial 
fractioning, leaving us with only one case. The technique
we employed to treat it is better illustrated by means of an 
example. Let us consider the two dimensional integral
\begin{equation}
I(\epsilon)=\int_{0}^{1}dx \int_{0}^{1}dy\, f(x,y)\,(1-y)^{-1+\epsilon}\,
	(1-x\,y)^{-1+\epsilon}\,,	
\end{equation}    
where $f(x,y)$ is a regular function in all the integration region. The 
integrand has poles along the curves $y=1$ and $y=1/x$ which intersect at
$x=1,\,y=1$. These singularities are regulated if $\epsilon>0$ (notice that
the integral remains finite even if the term $\epsilon$ in the exponent of 
$(1-x\,y)$ is absent). If one wrongly uses the recipe in 
eq.(\ref{eq:prescu1}) to prescribe the singularity in $y=1$ and then again
to deal with the pole in $x=1$ coming from the $\delta$ term, one ends with
ill-defined terms (more precisely terms with `plus' distributions which
are not integrable) and the leading singularity, in this case a double pole
$\epsilon^{-2}$, is accounted twice. The correct way to deal with this 
integral is to re-write it as
\begin{eqnarray}
I(\epsilon)&=&\int_{0}^{1}dx \int_{0}^{1}dy\, f(x,1)\,(1-y)^{-1+\epsilon}\,
	(1-x\,y)^{-1+\epsilon}\nonumber\\
	&&+\int_{0}^{1}dx \int_{0}^{1}dy\,\left(f(x,y)-f(x,1)\right)\,
	(1-y)^{-1+\epsilon}\,(1-x\,y)^{-1+\epsilon}\,.
\end{eqnarray}
The second term is integrable in the limit $\epsilon\rightarrow 0$ whereas in
the first one the integration over $y$ can be performed and gives
\begin{eqnarray}\label{eq:Ieps2}
I(\epsilon)&=&\int_{0}^{1}dx\,(1-x)^{-1+2\epsilon}\,f(x,1)\,
	\frac{_{2}F_{1}\left[
	\epsilon,2\,\epsilon,1+\epsilon;x\right]}{\epsilon}
	+\int_{0}^{1}dx \int_{0}^{1}dy\,\frac{f(x,y)-f(x,1)}
	{(1-y)(1-x\,y)}+{\cal O}(\epsilon)\,.
\end{eqnarray}
Now, the integral in the first term can be prescribed using (\ref{eq:prescu1}).
Doing that substitution, we end with the following identity:
\begin{eqnarray}
(1-y)^{-1+\epsilon}\,(1-x\,y)^{-1+\epsilon}&=&
	\left\{\frac{1}{2\,\epsilon^2}+\frac{\pi^2}{6}\right\}\,
	\delta(1-x)\,\delta(1-y)+
	\Biggl\{\frac{1}{\epsilon}\,
	\left(\frac{1}{1-x}\right)_{x[0,\underline{1}]}
	+2\,\left(\frac{\log(1-x)}{1-x}
	\right)_{x[0,\underline{1}]}\,\Biggr\}\,\delta(1-y)\nonumber\\
	&&+
	\left(\frac{1}{(1-y)(1-x\,y)}\right)_{y[0,\underline{1}]}
	+{\cal O}(\epsilon)\,.
\end{eqnarray}
The `plus' 
distribution $1/((1-y)(1-x\,y))_{y[0,\underline{1}]}$ stands for
the second term in the r.h.s. of eq.(\ref{eq:Ieps2}).
The factor $1/2$ in the double pole is a consequence of the fact that the
singular curve $y=1/x$ only intersects the integration region in a single
point. 

Prescriptions for all the singular (but regular at $u=1$) terms
appearing in the matrix elements
can be found, besides some 
subtleties related to the integration intervals in $\mbox{B1}$ and 
$\mbox{B2}$, with the technique shown in the example. 
Expressions for the prescriptions relevant in the $w=1$ region can be found in 
Appendix II.

The only remaining item is the prescription of singularities in $u=1$ in B2. 
These poles always appear as factors $1/(1-u)$ and only give rise to singular
integrals (when $\epsilon\rightarrow 0$) in terms proportional to $\delta(w)$ 
or $\delta(w_r-w)$ that come from the prescription of the singularities in the
$v$-$w$ plane. This is so because of the upper limit $w_r$ in the $w$ 
integration which goes to zero when $u\rightarrow 1$. For the $\delta$ terms,
the prescription of the singularities in $u=1$ can be done exactly as in
eq.(\ref{eq:prescu1}).


\section{Factorization of singularities}

As we mentioned in the previous section, once the angular integration 
and the prescription of the singularities in the $u$, $v$ and $w$ variables are
accomplished, the partonic cross sections exhibit a complex structure of poles
 in $\epsilon$. 
Explicit expressions for this structure in
region B0 can be found in Appendix III. Adding
virtual and real contributions all IR divergences cancel out, leaving us
with the UV and collinear poles. 
UV poles are canceled by means of coupling 
constant renormalization:
\begin{equation}
\frac{\alpha_{s}}{2\pi}=\frac{\alpha_s\left(M_R^2\right)}{2\pi}
	\left(1+\frac{\alpha_s\left(M_R^2\right)}{2\pi}\,f_{\Gamma}\,
	\frac{\beta_{0}}{\epsilon}
	\left(\frac{M_R^2}{4\pi\mu^2}\right)^
	{\epsilon/2}\right)\,,
\end{equation}
where $M_R$ is the renormalization scale and $\beta_0$ is the lowest order 
coefficient function in the QCD $\beta$ function:
\begin{equation}
\beta_{0}=\frac{11}{3}C_{A}-\frac{4}{3}n_f T_F\,
\end{equation}
with $C_A=N$ for $SU(N)$ and $T_F=1/2$ as usual; $n_f$ stands for the 
number of active quark flavours. 

Collinear singularities have to be factorized in the redefinition of 
parton densities, fragmentation and fracture functions. The redefinition
of parton densities is exactly the same as in totally inclusive DIS whereas
fragmentation functions are renormalized as they are in one-particle inclusive
electron-positron annihilation. Expressions for renormalized parton 
densities and fragmentation functions, up to order $\alpha_s^2$ and 
in the $\overline{MS}$ factorization scheme, can be found in
refs.\cite{zijli} and \cite{rijken} respectively. 

Notice that the renormalization of parton densities and fragmentation 
functions implies convolutions between the evolution kernels and the
SIDIS cross sections. At variance with the totally inclusive case, the
convolutions between the ${\cal O}(\alpha_s)$ cross section and 
the LO kernels include plus distributions in more than one variable which 
need to be handled with care. In order to make explicit the cancellations
between the ${\cal O}(\alpha_s^2)$ cross sections and these counterterms, 
the results of the above mentioned convolutions need to be expressed in 
terms of the very same variables used for the cross sections. 
One way to accomplish this is to retain to all orders in $\epsilon$ the 
${\cal O}(\alpha_s)$ cross sections, that is without replacing the 
singular factors like $(1-u)^{-1+\epsilon}$ in terms of distributions as 
described in the previous section,  
and rewrite the plus distributions in the LO kernels using
\begin{equation}
\left(\frac{1}{1-x}\right)_{+x[0,\underline{1}]}
	\rightarrow \lim_{\epsilon^{\prime}\rightarrow 0}\,\,
	(1-x)^{-1+\epsilon^{\prime}}-
	\frac{1}{\epsilon^{\prime}}\,\delta(1-x)+{\cal O}(\epsilon^{\prime})\,.
\end{equation}
In this way the appearance of plus distributions is avoided and the 
convolutions can be explicitly performed. The resulting expressions can 
be prescribed, keeping up to constant terms in $\epsilon$ and 
$\epsilon^{\prime}$, in exactly the same way as the ${\cal O}(\alpha_s^2)$ 
cross sections. Notice that at this point the poles in $\epsilon^{\prime}$ 
must cancel and the limit $\epsilon^{\prime}\rightarrow 0$ can be safely
taken, reflecting the fact that the LO kernels were already regular. The 
above mentioned procedure allows to extend the results of ref.\cite{arnold}
to SIDIS.

Once the renormalization of parton densities and fragmentation functions 
is accomplished, the remaining singularities occur in the region B0 
and are proportional to $\delta(1-w)$, that is the 
forward direction, so they have to be factorized into renormalized fracture 
functions. Otherwise, factorization would be broken. 
The bare fracture functions can be written in terms of renormalized
quantities as:
\begin{eqnarray}\label{eq:renfract}
M_{i,h/P}(\xi,\zeta)&=&
	\frac{1}{\xi}\int_{\xi}^{\frac{\xi}{\xi+\zeta}}\frac{du}{u}
	\int_{\frac{\zeta}{\xi}}^{\frac{1-u}{u}}\frac{dv}{v}\,
	\Delta_{ki\leftarrow j}(u,v,M_f)\,f_{j/P}^{r}\left(\frac{\xi}{u},
	M_f^2\right)
	\,D_{h/k}^{r}\left(\frac{\zeta}{\xi\,v},M_f^2\right)\nonumber\\
	&&+\int_{\frac{\xi}{1-\zeta}}^{1}\frac{du}{u}\,
	\Delta_{i\leftarrow j}(u,M_f)\,M_{j,h/P}^{r}
	\left(\frac{\xi}{u},\zeta,M_f^2\right)
\end{eqnarray}
where the factorization scale has been chosen to be the same for the three 
distributions. 
The functions  $\Delta_{i\leftarrow j}$ and  $\Delta_{ki\leftarrow j}$
are fixed in order to cancel all the remaining singularities 
in the cross section. 

The homogeneous kernels $\Delta_{i\leftarrow j}$ are the same that appear in the
inclusive case for parton densities and can be obtained from the corresponding
transition functions in ref.\cite{zijli}, whereas the non-homogeneous 
$\Delta_{ki\leftarrow j}$ are presented, for the case $j=g$, in this paper
for the first time. Explicitly:
\begin{eqnarray}\label{eq:renfractcon}
\Delta_{gg\leftarrow g}(u,v)&=&-\frac{\alpha_s}{2\pi}\,f_{\Gamma}\,\left(
	\frac{M_f^2}{4\pi\mu^2}\right)^{\epsilon/2}
	\frac{2}{\epsilon}\,\tilde{P}^{(0)}_{gg\leftarrow g}(u,v)\\
\Delta_{gq\leftarrow g}(u,v)&=&\Delta_{g\bar{q}\leftarrow g}(u,v)=
	\left(\frac{\alpha_s}{2\pi}\right)^{2}
	f_{\Gamma}^{\,2}\left(
	\frac{M_f^2}{4\pi\mu^2}\right)^{\epsilon}\Biggl\{
	\frac{2}{\epsilon^2}\biggl(\tilde{P}^{(0)}_{gq\leftarrow q}(u,v)
	\otimes P^{(0)}_{q\leftarrow g}(u)
	+\tilde{P}^{(0)}_{\bar{q}q\leftarrow g}(u,v)\otimes
	P^{(0)}_{g\leftarrow q}(v)\nonumber\\
	&&+\tilde{P}^{(0)}_{gg\leftarrow g}(u,v)
	\otimes^{\prime}P^{(0)}_{q\leftarrow g}(u)
	\biggr)
	-\frac{1}{\epsilon}P^{(1)}_{gq\leftarrow g}(u,v)
	\Biggr\}\\
\Delta_{\bar{q}q\leftarrow g}(u,v)&=&\Delta_{q\bar{q}\leftarrow g}(u,v)=
	-\frac{\alpha_s}{2\pi}\,f_{\Gamma}\,\left(
	\frac{M_f^2}{4\pi\mu^2}\right)^{\epsilon/2}
	\frac{2}{\epsilon}\,\tilde{P}^{(0)}_{\bar{q}q\leftarrow g}(u,v)
	+\left(\frac{\alpha_s}{2\pi}\right)^{2}
	f_{\Gamma}^{\,2}\left(
	\frac{M_f^2}{4\pi\mu^2}\right)^{\epsilon}\Biggl\{
	\frac{2}{\epsilon^2}\biggl(
	\tilde{P}^{(0)}_{\bar{q}q\leftarrow g}(u,v)
	\otimes P^{(0)}_{g\leftarrow g}(u)\nonumber\\
	&&+\tilde{P}^{(0)}_{\bar{q}q\leftarrow g}(u,v)\otimes
	P^{(0)}_{q\leftarrow q}(v)+\tilde{P}^{(0)}_{\bar{q}q\leftarrow g}(u,v)
	\otimes^{\prime}P^{(0)}_{q\leftarrow q}(u)
	+\frac{1}{2}\beta_{0}\,\tilde{P}^{(0)}_{\bar{q}q\leftarrow g}(u,v)
	\biggr)-\frac{1}{\epsilon}P^{(1)}_{\bar{q}q\leftarrow g}(u,v)
	\Biggr\}
\end{eqnarray}
where $\alpha_s$ is the bare coupling constant, the convolutions are defined as
\begin{eqnarray}\label{eq:conv}
f(u,v)\otimes g(u)&=&\int_{u}^{\frac{1}{1+v}}\,\frac{d\bar{u}}{\bar{u}}
	\,f(\bar{u},v)g\left(\frac{u}{\bar{u}}\right)\,,\nonumber\\
f(u,v)\otimes g(v)&=&\int_{v}^{\frac{1-u}{u}}\,\frac{d\bar{v}}{\bar{v}}
	\,f(u,\bar{v})g\left(\frac{v}{\bar{v}}\right)\,,\\
f(u,v)\otimes^{\prime} g(u)&=&\int_{u}^{1-u\,v}\,\frac{d\bar{u}}{\bar{u}}
	\,\frac{u}{\bar{u}}\,f\left(\bar{u},\frac{u}{\bar{u}}v\right)
	g\left(\frac{u}{\bar{u}}\right)\,,\nonumber
\end{eqnarray}
and  
\begin{equation}\label{eq:kertilde}
\tilde{P}^{(0)}_{ki\leftarrow j}(u,v)=P^{(0)}_{ki\leftarrow j}(u)\,
	\delta\left(v-\frac{1-u}{u}\right)\,.
\end{equation}
Finally the ${\cal O}(\alpha_s^2)$ kernels are given by
\begin{eqnarray}
P^{(1)}_{gq\leftarrow g}(u,v)&=&
 C_A\,T_F \,\Bigg\{-\frac{\left( 3 - 8\,u \right) \,u}{2} - 
\frac{4\,\left( 1 - u 
\right) \,u}{v} - \frac{8\,u^3}{{\left( 1 - u\,v \right) }^4} + 
  \frac{8\,u^2\,\left( 1 + u \right) }{{\left( 1 - u\,v \right) }^3}
 -   \frac{2\,u\,\left( 1 + 4\,u - 3\,u^2 \right) }
{{\left( 1 - u\,v \right) }^2}\nonumber\\
 &+&   \frac{2\,\left( 1 - 3\,u \right) \,u}{1 - u\,v}
+
\log\left(\frac{v}{1-u}\right)\frac{2\,\,
P^{(0)}_{q\leftarrow g}(u)}{v\,\left( 1 - u\,v \right) } +\log (1 + v)
\Bigg[ -u\,\left( 1 + 2\,u \right)  - u^2\,v + 
    \frac{2\,u\,P^{(0)}_{q\leftarrow g}(-u)}{1 - u\,v} \Bigg]\nonumber\\
&+&\log \left(\frac{1 - u - u\,v}{v}
	\right)
\Bigg[2\,\left( 1 - 3\,u \right) \,u 
-\frac{6\,P^{(0)}_{q\leftarrow g}(u)}{v} - 3\,u^2\,v  - 
+ \frac{2\,u\,\left( 1 + 4\,u \right) }{1 - u\,v}
  \frac{2\,u\,\left( 1 + 2\,u + 4\,u^2 \right) }{{\left( 1 - u\,v \right) }^2}
\nonumber\\
&+& 
  \frac{4\,u^2\,\left( 1 + u \right) }{{\left( 1 - u\,v \right) }^3}
- \frac{4\,u^3}{{\left( 1 - u\,v 
\right) }^4} 
 \Bigg]+\log(u)\,
\Bigg[ 4\,\left(-1 + u \right) \,u
+ 2\,u^2\,v + \frac{4\,P^{(0)}_{q\leftarrow g}(u)\,\left( 2 - u\,v \right)}
{v\,\left( 1 - u\,v \right) } \Bigg]
    \,
+ \log (1 - u\,v)\nonumber\\ 
&\times&\Bigg[u\,\left( 3 + 2\,u \right)  + \frac{2\,P^{(0)}_{q\leftarrow g}(u)}
{v} + u^2\,v + \frac{8\,u^3}{{\left( 1 - u\,v \right) }^4} - 
  \frac{8\,u^2\,\left( 1 + u \right) }{{\left( 1 - u\,v \right) }^3} 
 +\frac{4\,u\,\left( 1 + 2\,u + 4\,u^2 \right) }
{{\left( 1 - u\,v \right) }^2} - 
  \frac{4\,u\,\left( 1 + 3\,u \right) }{1 - u\,v} \Bigg]
\Bigg\} \nonumber\\
&+& C_F\,T_F \,\Bigg\{ 
4\,u + \frac{u^2\,\left( 1 - 4\,v \right) }{1 - u}
 -\frac{u^3\,v}{{\left( 1 - u \right) }^2} + 
  \frac{3}{{\left( 1 + v \right) }^2} - \frac{2 + 5\,u}{1 +
 v} 
+\log(v)\,\Bigg[\frac{u^3\,v}{{\left( 1 - u \right) }^2} + \frac{1}
{{\left( 1 + v \right) }^2} + 
  \frac{1 - 2\,u }{1 + v} \nonumber\\
&-& \frac{u^2\,\left( 2 + v \right) }{1 - u}  
\Bigg]+\log(1+v)\,\Bigg[ 
\frac{4\,P^{(0)}_{q\leftarrow g}(u)}{v} - \frac{2}{{\left( 1 + v \right) }^2} - \frac{2\,\left(
 1 - 2\,u \right) }{1 + v} + 2\,u^2\,\left( 2 + v \right)
\Bigg]
+\log(1-u)\Bigg[\frac{-4\,u^3}{1 - u} \nonumber\\
&+& \frac{2\,u^4\,v}{{\left( 1 - u 
\right) }^2} + 
  4\,P^{(0)}_{q\leftarrow g}(u)\,\left( \frac{4\,u}{1 - u} - \frac{3}{v} - \frac{2\,u^2\,v}
{{\left( 1 - u \right) }^2} \right) \Bigg]+\log(1-u-u\,v) 
 \Bigg[-2\,u - \frac{2\,u^3\,v}{{\left( 1 - u \right) }^2} - 
\frac{1}{{\left( 1 + v \right) }^2}\nonumber\\  
&-&  \frac{ 1 - 2\,u }{1 + v} 
+ \frac{2\,u^2\,\left( 2 + v 
\right) }{1 - u} + 
  6\,P^{(0)}_{q\leftarrow g}(u)\,\left(\frac{2}{v}- \frac{2\,u}{1 - u}   + \frac{u^2\,v}
{{\left( 1 - u \right) }^2} \right) \Bigg]
\Bigg\}\,,
\end{eqnarray}
and
\begin{eqnarray}
 P^{(1)}_{\bar{q}q\leftarrow g}(u,v)&=&
{C_A}\,T_{F}\Bigg\{- u\,\left( 1 - 4\,u \right)  - 
     {\frac{8}{u\,{{\left( 1 + v \right) }^4}}} + 
     {\frac{8\,\left( 1 + u \right) }{u\,{{\left( 1 + v \right) }^3}}} - 
     {\frac{2\,\left( 1 + 4\,u - 3\,{u^2} \right) }
       {u\,{{\left( 1 + v \right) }^2}}}
 + {\frac{2\,\left( 1 - 3\,u \right) }{1 + v}}+ 
      \log (1 - u\,v) \nonumber\\
&\times& \frac{\left( 1 - 2\,u\,v + {u^2}\,
	\left( 1 + {v^2} \right)  \right)}{1+v} + 
     \log (1 - u){\frac{\,P^{(0)}_{q\leftarrow g}(u)}{1 + v}}
 + \log \left({\frac{1 - u - u\,v}{1 - u}}\right)
	\Bigg[ {\frac{2}{1 + v}} - {\frac{3\,u}{1 - u - u\,v}} \Bigg] \,
      \,P^{(0)}_{q\leftarrow g}(u)\nonumber\\
&+& 
     \log (u)\,\Bigg[ 2\,\left( 1 - u \right) \,u + 2\,{u^2}\,v
- {\frac{2\,u\,P^{(0)}_{q\leftarrow g}(u)}{1 - u - u\,v}} \Bigg]  + 
     \log (1 + v)\,\Bigg[ u\,\left( 4 + u \right)  - 
        {{u}^2}\,v + {\frac{8}{u\,{{\left( 1 + v \right) }^4}}} - 
        {\frac{8\,\left( 1 + u \right) }{u\,{{\left( 1 + v \right) }^3}}} 
\nonumber\\
&+& {\frac{4\,\left( 1 + 2\,u + 4\,{u^2} \right) }
          {u\,{{\left( 1 + v \right) }^2}}} - 
        {\frac{4\,\left( 1 + 3\,u \right) }{1 + v}} + 
        {\frac{2\,u\,P^{(0)}_{q\leftarrow g}(u)}{1 - u - u\,v}} \Bigg]  + 
     \log \left({\frac{1 - u - u\,v}{v}}\right)\,
      \Bigg[ u\,\left( 1 + 3\,u \right) 
- 3\,{u^2}\,v 
\nonumber\\
&+& 
        {\frac{4}{u\,{{\left( 1 + v \right) }^4}}}
- 
        {\frac{4\,\left( 1 + u \right) }{u\,{{\left( 1 + v \right) }^3}}} + 
        {\frac{2\,\left( 1 + 2\,u + 4\,{u^2} \right) }
          {u\,{{\left( 1 + v \right) }^2}}} - 
        {\frac{2\,\left( 1 + 4\,u \right) }{1 + v}} + 
        {\frac{3\,u\,P^{(0)}_{q\leftarrow g}(u)}{1 - u - u\,v}} \Bigg]
  -4\,\Big[u\,\left( 1 - u \right)\nonumber\\
 &+&\log (a_{f})\,P^{(0)}_{q\leftarrow g}(u) \Big] \,
      {{\left({\frac{1}{a_{f} - v}}\right)}_{v\left[0,\underline{a_{f}}
\right]}}+ 4\,P^{(0)}_{q\leftarrow g}(u)\,
      {{\left({\frac{\log (a_{f} - v)}{a_{f} - v}}\right)}_
        {v\left[0,\underline{a_{f}}\right]}}
	+ \Big[ u - 4\,\left( 1 - u \right) \,u\,\log (1 - u)\nonumber\\
&-& 
        \Big( {{\log (1 - u)}^2} + 2\,\log (1 - u)\,\log (u)
+ 2\,Li_{2}(u) \Big) \,P^{(0)}_{q\leftarrow g}(u)
        \Big] \,\delta (a_{f} - v) \Bigg\}  + 
 C_F\, T_{F}\,\Bigg\{ {\frac{4\,{u^2}\,v}{1 - u}} + 
     {\frac{{u^3}\,v}{{{\left( 1 - u \right) }^2}}} \nonumber\\
&+& 
     {\frac{3\,{u^2}}{{{\left( 1 - u\,v \right) }^2}}} 
- {\frac{2\,u + 5\,{u^2}}{1 - u\,v}} + 
     \log (1 - u - u\,v)
\Bigg[ -{\frac{{u^3}\,v}{{{\left( 1 - u \right) }^2}}}
 + {\frac{{u^2}}{{{\left( 1 - u\,v \right) }^2}}} + 
        {\frac{\left( 1 - 2\,u \right) \,u}{1 - u\,v}}
- {\frac{u\,\left( 1 - u\,v \right) }{1 - u}} \Bigg] \nonumber\\
&+& 2\,\log (1 - u\,v)\,
      \Bigg[ u\,\left( 1 + u \right)  - {u^2}\,v - 
        {\frac{{u^2}}{{{\left( 1 - u\,v \right) }^2}}} - 
        {\frac{\left( 1 - 2\,u \right) \,u}{1 - u\,v}} 
+ {\frac{u\,P^{(0)}_{q\leftarrow g}(u)}{1 - u - u\,v}} \Bigg]  + 
    4 \log (u)\,\Bigg[ -u\,\,\left( 1 - u \right)\nonumber\\  
&-& {u^2}\,v + 
        {\frac{u\,P^{(0)}_{q\leftarrow g}(u)}{1 - u - u\,v}} \Bigg]  + 
     \log (v)\,\Bigg[ {\frac{2\,{u^2(1-v)}}{1 - u}}
       + {\frac{2\,{u^3}\,v}{{{\left( 1 - u \right) }^2}}}- 
        {\frac{{u^2}}{{{\left( 1 - u\,v \right) }^2}}} - 
        {\frac{\left( 1 - 2\,u \right) \,u}{1 - u\,v}}\nonumber\\ 
&+&    {\frac{6\,{u^3}\,{v^2}\,P^{(0)}_{q\leftarrow g}(u)}
          {{{\left( 1 - u \right) }^2}\,\left( 1 - u - u\,v \right) }} \Bigg]
        +2\, \log (1 - u)\,\Bigg[  
        \left( {\frac{4\,u}{1 - u}}- {\frac{3\,u}{1 - u - u\,v}} 
          + {\frac{4\,{u^2}\,v}{{{\left( 1 - u \right) }^2}}}
 \right) \,P^{(0)}_{q\leftarrow g}(u)-{\frac{{u^4}\,v}
{{{\left( 1 - u \right) }^2}}} \nonumber\\
&-&  {\frac{{{u}^3}}{(1 - u)}}
        \Bigg]  
+ 4\,\Bigg[ -\frac{u}{4} + \left( 1 - u \right) \,u\,
         \log\left(a_f\right) + 
        \left( \zeta(2) + \frac{{\log (1 - u)}^2}{4} + 
           Li_{2}(u) \right) \,P^{(0)}_{q\leftarrow g}(u)
        \Bigg] 
	\,\delta \left(a_f - v\right) \Bigg\}
\end{eqnarray}
where $a_f=(1-u)/u$. Although $\Delta_{qg\leftarrow g}$ is formally a 
NLO kernel, it occurs 
for the first time at order $\alpha_s^3$ thus it does not show up in the 
present calculation. Notice that the NLO kernels depend on both $u$ and $v$ 
variables and that this dependence cannot be factorized.

Once obtained the explicit expressions for the relations between renormalized
and bare fracture functions, we can easily derive the evolution equations
for the renormalized fracture functions, which can be written as
\begin{eqnarray}\label{eq:fractev}
&&\frac{\partial \,M^{r}_{i,h/P}(\xi,\zeta,M^2)}{\partial \log M^2}=
	\frac{\alpha_s(M^2)}{2\pi}
	\int_{\frac{\xi}{1-\zeta}}^{1}\frac{du}{u}\,
	\left[
	\,P^{(0)}_{i\leftarrow j}(u)
	+\frac{\alpha_s(M^2)}{2\pi} 
	\,P^{(1)}_{i\leftarrow j}(u)\right]
	\,M_{j,h/P}^{r}
	\left(\frac{\xi}{u},\zeta,M^2\right)\nonumber\\
	&&+\frac{\alpha_s(M^2)}{2\pi}\,\frac{1}{\xi}
	\int_{\xi}^{\frac{\xi}{\xi+\zeta}}\frac{du}{u}
	\int_{\frac{\zeta}{\xi}}^{\frac{1-u}{u}}\frac{dv}{v}\,
	\left[
	\tilde{P}^{(0)}_{ki\leftarrow j}(u,v)+\frac{\alpha_s(M^2)}{2\pi}\,
	P^{(1)}_{ki\leftarrow j}(u,v)\right]
	\,f_{j/P}^{r}\left(\frac{\xi}{u},
	M^2\right)
	\,D_{h/k}^{r}\left(\frac{\zeta}{\xi\,v},M^2\right)\,,
\end{eqnarray}
where the NLO kernels $P^{(1)}_{i\leftarrow j}(u)$ are $1/8$ of those given in 
ref.\cite{hamberg} due to the different conventions implemented. At variance 
with the LO case where the kernels are 
proportional to $\delta(v-(1-u)/u)$, the NLO kernels have support
in all the integration region in the non homogeneous term of 
eq.(\ref{eq:fractev}). Due to this fact, at NLO, the non homogeneous terms in 
the evolution equations do not take the familiar form given in eq.(12) of 
ref.\cite{ven}. In terms of moments, eq.(\ref{eq:fractev}) can be written as:
\begin{eqnarray}
\frac{\partial \,M^{r}_{i,h/P}[m,n]}{\partial \log M^2}&=&
	M^{r}_{i,h/P}[m,n]\,P_{i\leftarrow j}[m]+f_{j/P}^{r}[m+n-1]
	\,D_{h/k}^{r}[n]\,\hat{P}_{ki\leftarrow j}[m-1,n]\,
\end{eqnarray}
where the moments are defined as
\begin{eqnarray}
F[m,n]&=&\int_{0}^{1}\frac{d\xi}{\xi}\,\int_{0}^{1-\xi}\frac{d\zeta}{\zeta}
	\,\xi^{m}\,\zeta^{n}\,F(\xi,\zeta)\,,\,\,\,\,
F[m]=\int_{0}^{1}\frac{d\xi}{\xi}\,\xi^{m}\,F(\xi)\,,
\end{eqnarray}
and
\begin{equation}
\hat{P}_{ki\leftarrow j}[m,n]=\int_{0}^{1}\frac{d\xi}{\xi}\,\int_{0}^{1-\xi}
\frac{d\zeta}{\zeta}
	\,\xi^{m}\,\zeta^{n}\,P_{ki\leftarrow j}\left(\xi,
\frac{\zeta}{\xi}\right)
\end{equation}

\begin{figure}[t] 
\begin{center}
\begin{minipage}{55mm}
\begin{center}
\includegraphics[height=5.0cm,width=5.0cm]
{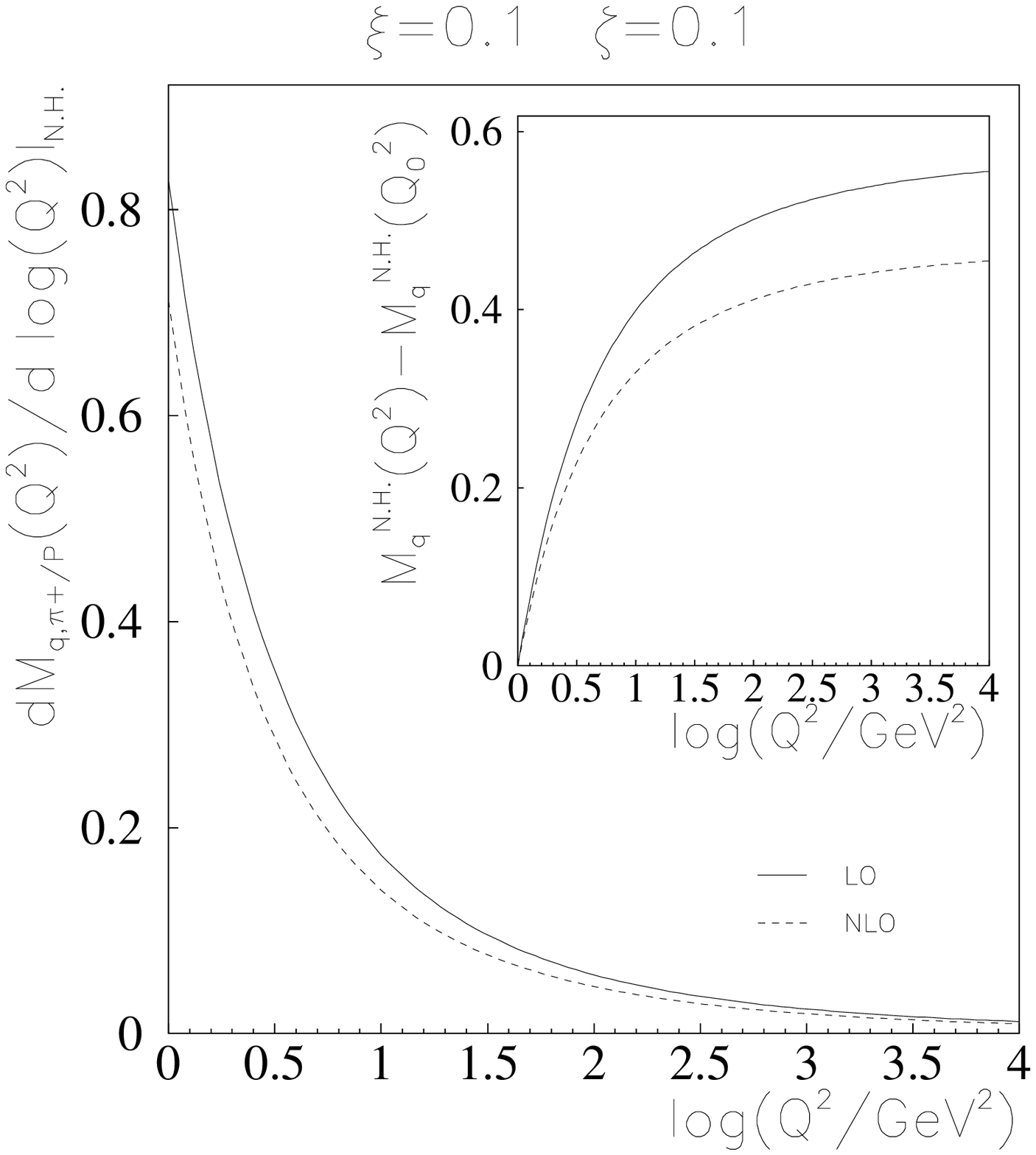}
\end{center}
\end{minipage}
\begin{minipage}{55mm}
\begin{center}
\includegraphics[height=5.0cm,width=5.0cm]
{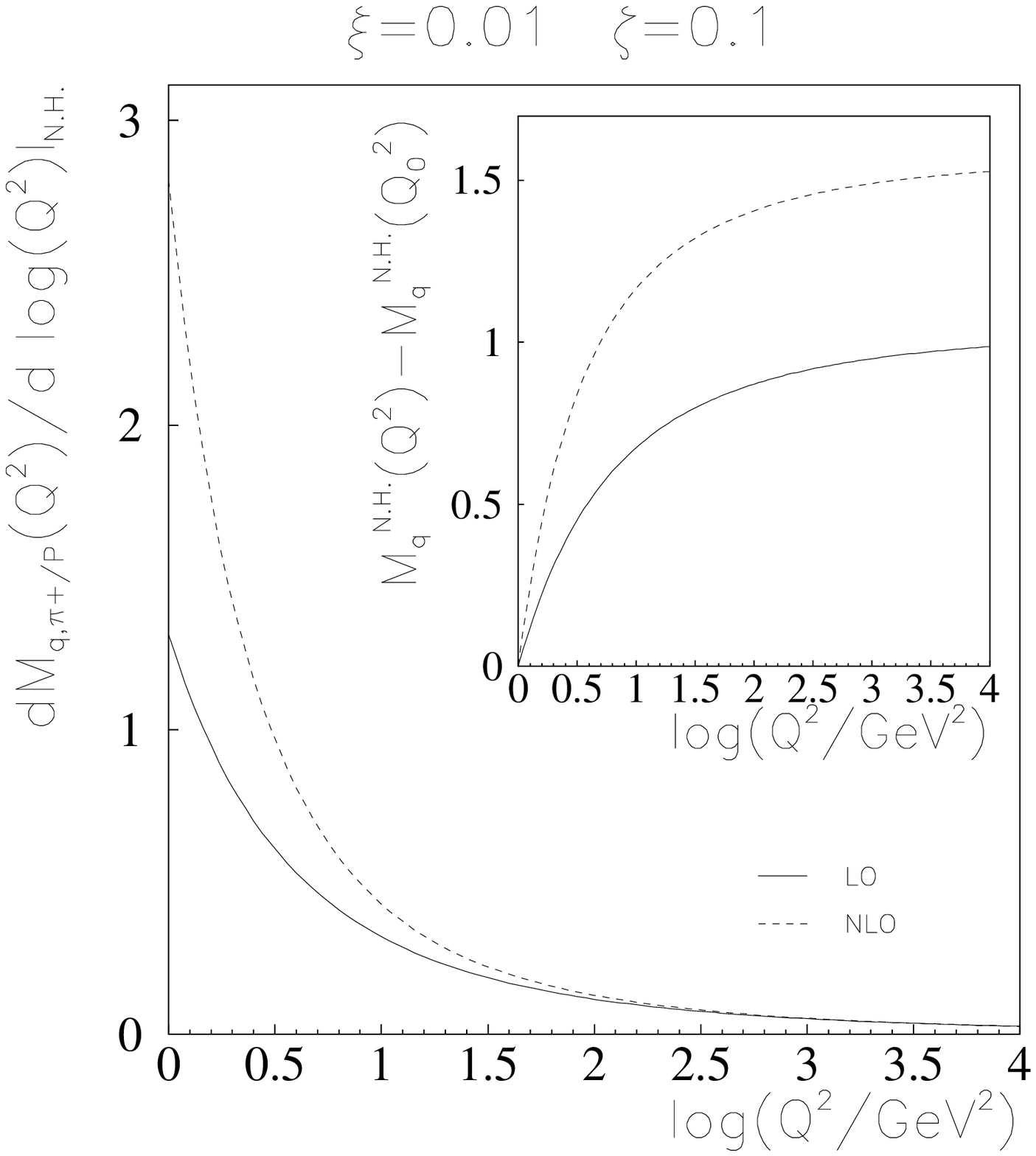}
\end{center}
\end{minipage}
\begin{minipage}{55mm}
\begin{center}
\includegraphics[height=5.0cm,width=5.0cm]
{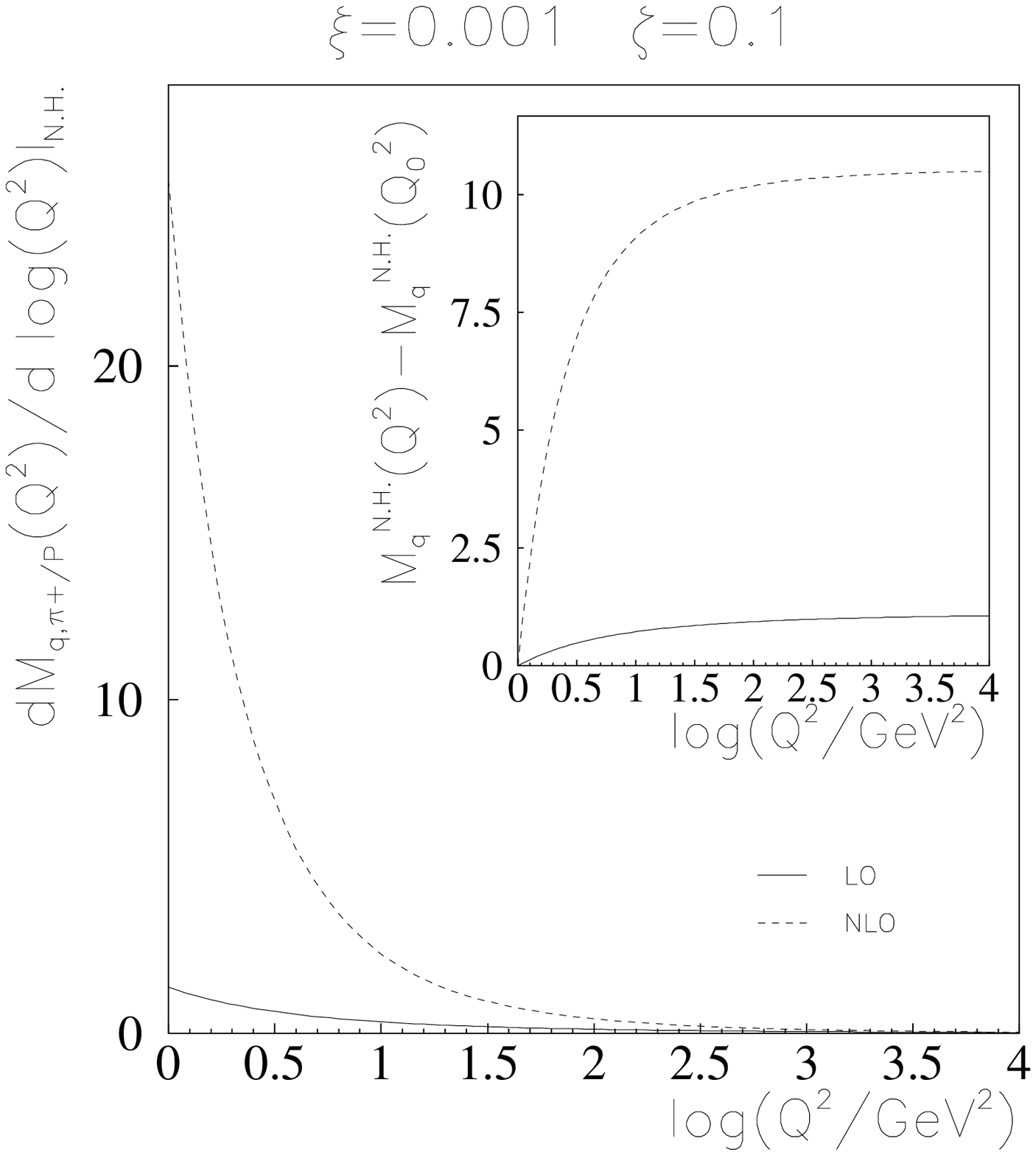}
\end{center}
\end{minipage}
\begin{minipage}{55mm}
\begin{center}
\includegraphics[height=5.0cm,width=5.0cm]
{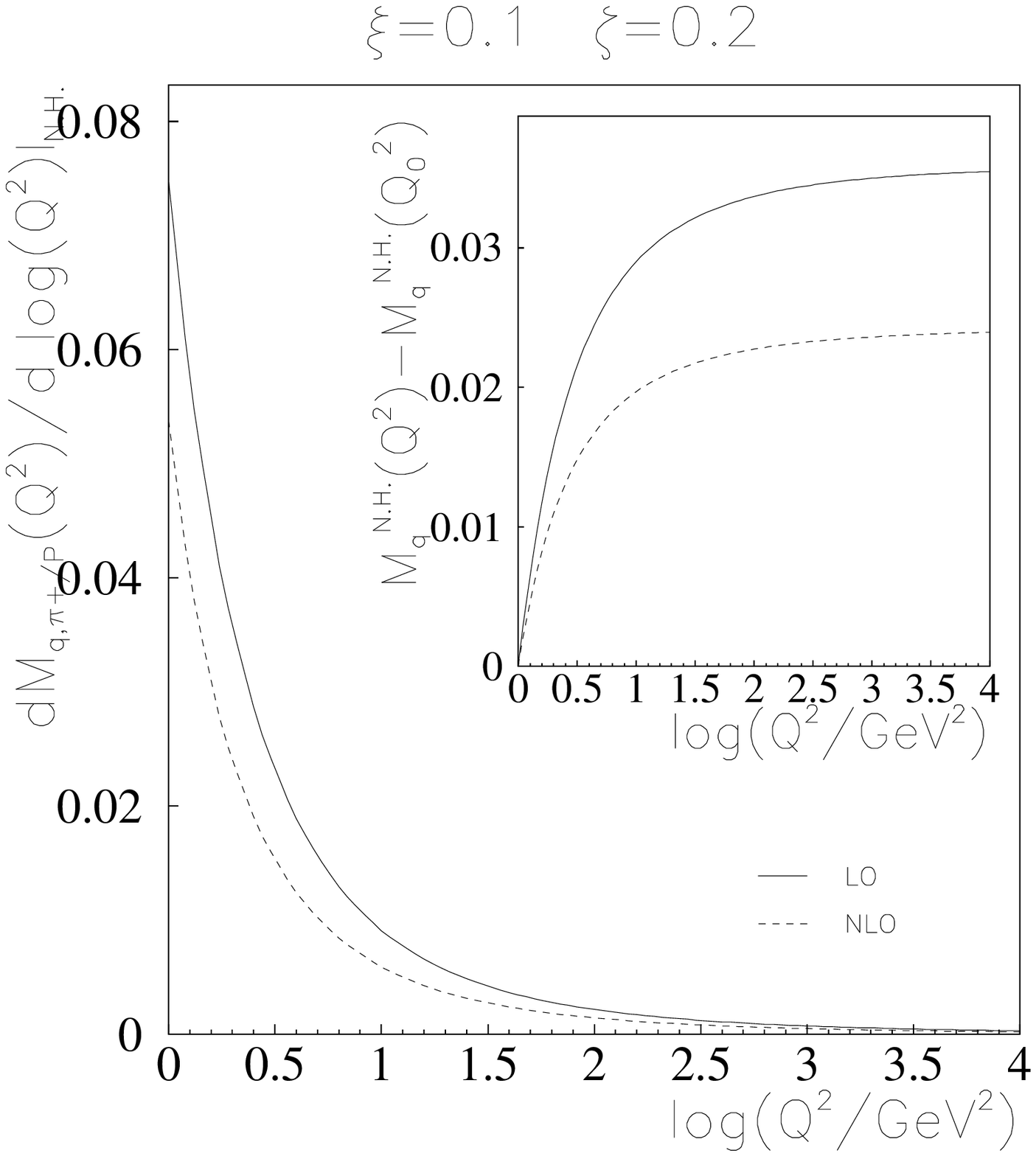}
\end{center}
\end{minipage}
\begin{minipage}{55mm}
\begin{center}
\includegraphics[height=5.0cm,width=5.0cm]
{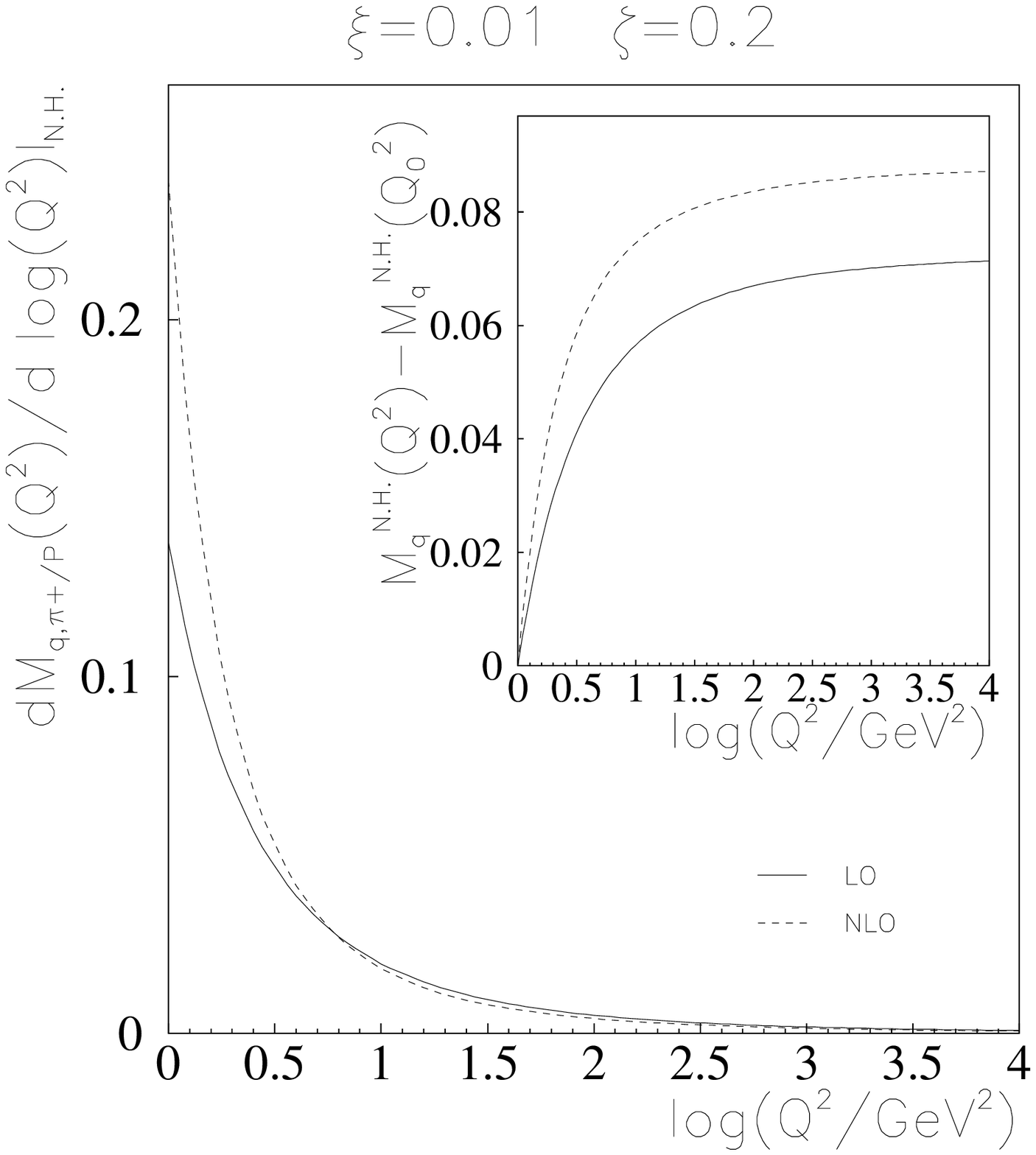}
\end{center}
\end{minipage}
\begin{minipage}{55mm}
\begin{center}
\includegraphics[height=5.0cm,width=5.0cm]
{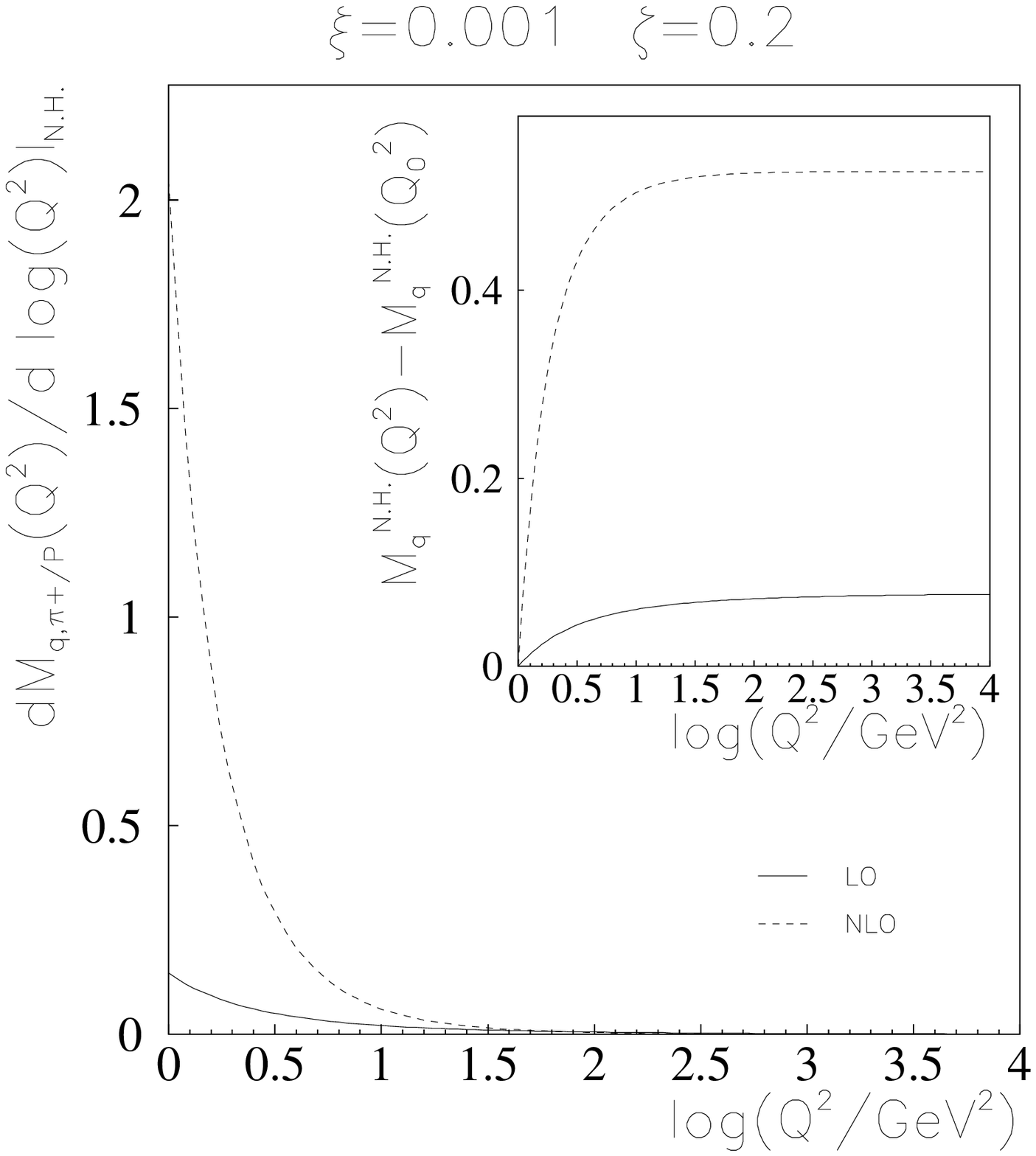}
\end{center}
\end{minipage}
\end{center}
\caption{Non homogeneous contributions to the derivative of 
$M_{q}$ for different values of $\xi$ and $\zeta$. Inset plots
show the integral over $Q^2$ of this contributions taking 
$M_{q}^{N.H.}(Q_{0}^2)=0$ with $Q_{0}=1\,\mbox{GeV}$ as a reference.}
\label{fig:numcomp}
\end{figure}  
Figure \ref{fig:numcomp} compares (for different
values of $\xi$ and $\zeta$) the relative
size of the LO and NLO contributions to the non homogeneous term in the 
evolution equation (\ref{eq:fractev})
computed with standard sets of parton distributions \cite{GRV} and 
fragmentation functions \cite{kretzer} for the case $i=q$ and $h=\pi^{+}$.
The inset plots show the integral over $Q^2$ of this contributions. 
Notice that only those terms proportional to $f_{g/P}$ were taken into 
account in the ${\cal O}(\alpha_s^2)$ pieces. 
In reference \cite{prd97} it was found that the
LO non homogeneous contribution falls rapidly as $\zeta$ grows.
This behavior is related to the shrinkage of the integration region and with
the fall of fragmentation functions, $D_{h/i}(z)$, in the limit 
$z\rightarrow 1$.
This is also the case of the NLO contributions. At moderate and large values 
of $\xi$ ($\xi\ge 0.1$)
the ${\cal O}(\alpha_s^2)$ contributions are typically one order of magnitude 
smaller than the ${\cal O}(\alpha_s)$
ones so NLO and LO results differ only by a few percents. 
This can be traced back to the extra power of $\alpha_s$ and the 
small size of the integration region since the 
interval of the $v$ integral in eq.(\ref{eq:fractev}) shrinks to 
the point $(1-u)/u$ when $\xi\rightarrow 1-\zeta$. However, when $\xi$
diminishes the integration region expands and NLO contributions grow 
considerably faster than the LO ones which are kinematically restricted to 
the curve $v=(1-u)/u$. The remarkable growth of the ${\cal O}(\alpha_s^2)$ 
terms makes these contributions even larger than the constrained 
${\cal O}(\alpha_s)$ pieces at lower values of $\xi$ and thus a priori 
non negligible in the evolution equations.

Of course, in order to assess the actual relevance of the NLO 
non homogeneous effects in the full evolution of fracture functions, one needs
a realistic (based on actual data) estimate for the size and shape for 
these functions at a given scale,
and compute the evolution taking into account all the appropriate kernels, but
our present results suggest that non homogeneous NLO effects could be relevant.

\section{Summary and conclusions}

In this paper we have computed the  ${\cal O}(\alpha_s^2)$ gluon initiated 
QCD corrections to one particle inclusive deep inelastic processes. At 
variance with the inclusive case, in one particle inclusive processes 
the kinematical characterization of the final state particle requires to 
preserve the full dependence of the amplitude in the relevant variables. 
This impedes the cancellation of some singularities to be later factorized 
into fracture functions and leads to a non trivial singularity structure.  
In order to deal with this we have highlighted the importance of collecting 
to all orders the potentially singular factors in the 3-particle final state 
angular integrals and implemented a general approach for the 
prescription of overlapping singularities.

By the explicit replacement of the bare parton densities, fragmentation and
fracture functions with the corresponding renormalized quantities in both the
${\cal O}(\alpha_s)$ and ${\cal O}(\alpha_s^2)$ cross sections, we have 
explicitly verified the factorization of collinear singularities 
obtaining for the first time the relevant kernels at this order. 
In doing so, we give a recipe for dealing with convolutions of distributions 
in more than one variable which occur in the
computation of the $\alpha_s^2$ contribution coming from the convolution of 
${\cal O}(\alpha_s)$ cross sections and renormalized functions.
We also
derived the evolution equations for fracture functions valid at NLO.

Regarding the phenomenological consequences of these corrections, we have 
found that the ${\cal O}(\alpha_s^2)$ contributions to the evolution 
equations are mild in most of the kinematical range, however they are as 
important or even larger than the
$\alpha_s$ ones for small values of $x_B$, where these last contributions are 
suppressed by the available phase space. This behaviour, at variance with the
LO case, allows the non homogeneous effects to be sizeable even at larger
hadron momentum fractions, thus being relevant for the scale dependence for
diffractive and leading baryon processes.  

\begin{acknowledgments}
We are indebted to D. de Florian for valuable
comments and discussions. We also acknowledge S. Catani and W. L. van Neerven 
for their generous help. CAGC thanks J. Bernabeu and the Departamento de 
F\'{\i}sica Te\'orica, Universidad de Valencia for the warm hospitality 
extended to him.
\end{acknowledgments}

\section*{Appendix I}

In this appendix we present the results obtained for the
coefficients $D^{(1)}$ in the
${\cal O}(\alpha_s)$ cross sections. They are
\begin{eqnarray}
D_{1qq,M}^{(1)}&=&\frac{C_{F}}{2}\,\Bigg\{ 
     \Bigg[ {{\left( 1 - u \right) \,\log\left({\frac{1 - u}{u}}\right)}} + 
       {\left( {\frac{{{\pi }^2}}{6}} + 
             {\frac{{{\log \left({\frac{1 - u}{u}}\right)}^2}}{2}} \right) \,
           {p_{gq\leftarrow q}}(u)} \Bigg]\,\delta (1 - v)\,\delta (w)+ 
    \delta (w_{r}-w) \nonumber\\
&\times&\Bigg[ 
       {{p_{gq\leftarrow q}}(u)}\,
{{\left({\frac{\log (1 - v)}{1 - v}}\right)}_{+v[a,\underline{1}]}} + 
       {\left( 1 - u +   \log \left({\frac{1 - u}{(1-a)\,u}}
\right) \,{p_{gq\leftarrow q}}(u) \right)\,
{{\left({\frac{1}{1 - v}}\right)}_{+v[a,\underline{1}]}} }\nonumber\\ 
&-&{\frac{2\,(v-a)}{{{\left( 1 - a \right) }^2}\,\left( 1 - u \right) }} + 
  {\frac{\left( 1 -v- 2\,\left( 1 - a \right) \,u \right) \,
       }{{{\left( 1 - a \right) }^2}\,\left( 1 - u \right) }} 
\left[\log\left({\frac{\left( 1 - u \right) \,\left( 1 - v \right) }
           {\left( 1 - a \right) \,u}}\right) + 
\log \left({\frac{v-a}{1 - a}}\right)
        -1 \right]\nonumber\\
&+& \frac{1}{1 - v }
\log \left({\frac{v-a}{1 - a}}\right)\,{p_{gq\leftarrow q}}(u)
\Bigg] 
     \Bigg\}\,,\\
\nonumber\\
D_{1qg,M}^{(1)}&=&\frac{C_F}{2}\Bigg\{ 
       \Bigg[ \left( 1 - u \right) \,\log\left({\frac{1 - u}{u}}\right) + 
          \left( {\frac{{{\pi}^2}}{6}} +\frac{ 
                {{\log\left({\frac{1 - u}{u}}\right)}^2}}{2} \right) \,
{p_{{gq\leftarrow q}}(u)}
               \Bigg] \,\delta (v-a)\,\delta (1 - w)
+ \delta(w_r-w)\nonumber\\
&\times&\Bigg[  {p_{{gq\leftarrow q}}(u)}\,\left({{\frac{\log (v-a)}
             {v-a}}}\right)_{+v[\underline{a},1]}+
     \left( 1 - u + \log\left({\frac{1 - u}{\left( 1 - a \right) \,u}}
\right)\,
           {{{p}}_{{gq\leftarrow q}}(u)} \right) \,
        {\left({{\frac{1}{v-a}}}\right)_{+v[\underline{a},1]}}\nonumber\\ 
&-&{\frac{2\,\left( 1 - v \right) }
         {{{\left( 1 - a \right) }^2}\,\left( 1 - u \right) }} + 
       {\frac{\left(v -a - 2\,\left( 1 - a \right) \,u \right) \,
            }{{{\left( 1 - a \right) }^
             2}\,\left( 1 - u \right) }}
\left[\log \left({\frac{\left( 1 - u \right) \,
                  \left( 1 - v \right) }{\left( 1 - a \right) \,u}}\right) + 
             \log \left({\frac{v-a}{1 - a}}\right)-1 \right]\nonumber\\
&+&       \frac{1}{v-a}\log\left({\frac{1 - v}{1 - a}}\right)\,
           {{{p}}_{{gq\leftarrow q}}(u)}\Bigg]  \Bigg\}\,,\\
\nonumber\\
D_{1gq,M}^{(1)}&=& 
\frac{{T_F}}{2}\,\Bigg\{ \Bigg[ {\frac{{{\pi }^2}}{6}}+ 
          2\,\left( 1 - u \right) \,u\,\left(
\log\left({\frac{1 - u}{u}}\right)-1-\frac{\pi^2}{6}\right) + 
          {\frac{{{\log\left({\frac{1 - u}{u}}\right)}^2}\,
{p_{\overline{q}q\leftarrow g}}(u)}{2}}
           \Bigg] \,\big( \delta (v-a)\,\delta (1 - w)  \nonumber\\
&+& 
          \delta (1 - v)\,\delta (w) \big)+ 
       \Bigg[ -{\frac{2\,\left( 1 + a - 2\,v \right) \,\log (1 - a)}
            {{{\left( 1 - a \right) }^2}}} - 
          {\frac{2}{1 - a}}\,\bigg( 
                \log\left({\frac{1 - u}{\left( 1 - a \right) \,u}}\right) + 
                \log\left({\frac{1 - v}{1 - a}}\right) \nonumber\\ 
 &+& 
	\log\left({\frac{v-a}{1 - a}}\right)-1\bigg)+
          \left( {\frac{1}{v-a}\log\left({\frac{1 - v}{1 - a}}\right)} + 
             {\frac{1}{1 - v}\log\left({\frac{v-a}{1 - a}}\right)} \right) \,
           {p_{\overline{q}q\leftarrow g}}(u) + 
          \bigg( 2\,\left( 1 - u \right) \,u  \nonumber\\
&+& 
             \log \left({\frac{1 - u}{\left( 1 - a \right) \,u}}\right)\,
              {{{p}}_{{\overline{q}q\leftarrow g}}(u)} \bigg)
       \bigg( {\left({{\frac{1}{1 - v}}}\right)_{+v[a,\underline{1}]}} 
+{\left({{\frac{1}{v - a}}}\right)_{+v[\underline{a},1]}}\bigg) 
          +{p_{\overline{q}q\leftarrow g}}(u)\,
           \bigg( {\left({{\frac{\log (1 - v)}{1 - v}}}\right)_
	{+v[a,\underline{1}]}}  \nonumber\\
&+& 
             {\left({{\frac{\log (-a + v)}{v-a}}}\right)_
               {+v[\underline{a},1]}} \bigg)  \Bigg]\,\delta (w_{r}-w)
        \Bigg\}\,,\\
\nonumber\\ 
D_{2qq,M}^{(1)}&=&
{C_F}\,\Bigg\{ \Bigg[- {\frac{2\,u\,
            \left( 1 - {x_B} \right)  + {x_B}}{u - 
           {x_B}}} - {\frac{1 + u}{2\,\left( 1 - v \right)}}
\,\log\left({\frac{v-a}{1 - a}}\right) 
 +{\frac{1 + v}{2\,
           \left( 1 - u \right) }{ \,
           \log\left({\frac{u - {x_B}}{1 - {x_B}}}\right)}}\nonumber\\
&+&        \frac{1}{2}\,\bigg( {\frac{2\,{u^2}\,\left( 1 - {x_B} \right) }
               {{{\left( u - {x_B} \right) }^2}}} - 
             {\frac{u\,\left( 1 + v \right) \,
                 \left( 1 - {x_B} \right) \,{x_B}}{{{\left
                     ( u - {x_B} \right) }^2}}}
 -            {\frac{\left( 1 + v \right) \,{x_B}}
               {u - {x_B}}} \bigg)
{ \,\left( 1 + \log\left({\frac{\left( 1 - u \right) \,
                  \left( 1 - v \right) }{\left( 1 - a \right) \,v}}\right) + 
             \log\left({\frac{v-a}{1 - a}}\right) \right) }\nonumber\\ 
&+&       {\frac{1 + {v^2}}{2\,
           \left( 1 - u \right) \,\left( 1 - v \right) }} \,
           {\log\left({\frac{v-a}{\left( 1 - a \right) \,v}}\right)}
- {\frac{1 + v}{2} \,{
           {\left({{\frac{\log (1 - u)}{1 - u}}}\right)_
{+u[0,\underline{1}]}}}}
-  {\frac{1 + u}{2} \,{{\left({{\frac{\log (1 - v)}{1 - v}}}\right)
_{+v[0,\underline{1}]}}}}
\nonumber\\
&+& \bigg( {\frac{1 - u}{2}} 
-          {\frac{\left( 1 + u \right) \,\log (1 - u)}{2}} - 
          {\frac{\left( 1 + {u^2} \right)}{2\,
              \left( 1 - u \right) } \,
              \log\left({\frac{u - {x_B}}{1 - {x_B}}}\right)}
+{\left({{\frac{\log (1 - u)}{1 - u}}}\right)_{+u[0,\underline{1}]}}
 \bigg) 
	\,{\left({{\frac{1}{1 - v}}}\right)_{+v[0,\underline{1}]}}\nonumber\\
&+&        
       \bigg( {\frac{1 - v}{2}}
-           {\frac{\left( 1 + v \right) \,\log (1 - v)}{2}} + 
          {\frac{\left( 1 + {v^2} \right) \,\log (v)}
            {2\,\left( 1 - v \right) }}+
{\left({{\frac{\log (1 - v)}{1 - v}}}\right)_{+v[0,\underline{1}]}}
 \bigg) \,
        {\left({{\frac{1}{1 - u}}}\right)_{+u[0,\underline{1}]}}      
 \Bigg] \,\delta (w_r-w) \nonumber\\
&+& \Bigg[ \Bigg( -{\frac{{{\pi }^2}}{12}} \left( 1 + v \right)+ 
          {\frac{\log (1 - v)\,\log (v)}{1 - v}} + 
          {\frac{{{\log (v)}^2}}{2\,\left( 1 - v \right) }} + 
          {\frac{\left( 1 - v \right) \,\log (\left( 1 - v \right) \,v)}
            {2}} - {\frac{\left( 1 + v \right) \,
              {{\log (\left( 1 - v \right) \,v)}^2}}{4}}\nonumber\\ 
&+&           {\frac{{\pi }^2}{6}\,
{{\left({{\frac{1}{1 - v}}}\right)_{+v[0,\underline{1}]}}}} + 
          {\frac{1}{2}
{{\left({{\frac{{{\log (1 - v)}^2}}{1 - v}}}\right)_{+v[0,\underline{1}]}}}} 
\Bigg) \,\delta (1 - u) + \Bigg( -{\frac{{\pi }^2}{12}} 
\,                \left( 1 + u \right)  - 
          {\frac{\log (1 - u)}{1 - u}}\,\log\left({\frac{u - {x_B}}
                 {1 - {x_B}}}\right)\nonumber\\ 
&+&           {\frac{1}{2\,\left( 1 - u \right) }}
\,{{\log\left({\frac{u - {x_B}}{1 - {x_B}}}\right)}^2} + 
          {\frac{\left( 1 - u \right) }{2}}
\,              \log\left({\frac{1-u}
                 {\left( 1 - a \right) \,u}}
               \right)
 + {\frac{\left( 1 + u \right)}{4}}
\,              {{\log\left({\frac{1-u}
                 {\left( 1 - a \right) \,u}}
               \right)}^2}+
{\frac{{{\pi }^2}}{6}}\,
{\left({{\frac{1}{1 - u}}}\right)_{+u[0,\underline{1}]}}\nonumber\\
 &+&          {\frac{1}{2}}\,
{\left({{\frac{{{\log (1 - u)}^2}}{1 - u}}}\right)_{+u[0,\underline{1}]}} 
\Bigg) 
         \,\delta (1 - v) + \delta (1 - u)\,\delta (1 - v)\,
        \left( 8 + {\frac{{{\pi }^2}}{4}} - 2\,\zeta (3) \right)
\Bigg]\delta(w) 
     \Bigg\}\,,\\
\nonumber\\
D_{2qg,M}^{(1)}&=&
\frac{C_F}{2}\,\Bigg\{ 
        \Bigg[ v\,\log (\left( 1 - v \right) \,v) +\left( 
          {\frac{{\pi }^2}{6}} + 
          {\frac{{{\log (\left( 1 - v \right) \,v)}^2}}{2}}\right)
\,              {p_{g\leftarrow q}}(v) \Bigg]\,\delta (1 - u)\,\delta (w)  
+ \delta (w_r-w)\nonumber\\
&\times&        \Bigg[
           {p_{g\leftarrow q}}(v)\, 
          {\left({{\frac{\log (1 - u)}{1 - u}}}\right)_{u[0,\underline{1}]}}
+         \left( v - \log (\left( 1 - v \right) \,v)\,
              {p_{g\leftarrow q}}(v) \right)
{\left({{\frac{1}{1 - u}}}\right)_{u[0,\underline{1}]}}
+{\frac{\left( 1 - u \right) \,{u^2}}
            {\left(v-a \right) \,
              {{\left( u - {x_{B}} \right) }^2}}}\nonumber\\ 
&-&           {\frac{2\,{u^2}\,v\,\left( 1 - {x_{B}} \right) }
            {\left( v-a \right) \,
              {{\left( u - {x_{B}} \right) }^2}}} + 
          {\frac{v\,{x_{B}}}
            {u\,\left( v-a \right) \,\left( 1 - {x_{B}} \right) }}
           + {\frac{u\,{v^2}\,\left( 1 - {x_{B}} \right) \,
              {x_{B}}}{\left( v-a \right) \,
              {{\left( u - {x_{B}} \right) }^2}}} + 
          {\frac{{v^2}\,{x_{B}}}
            {\left( v-a \right) \,\left( u - {x_{B}} \right) }}\nonumber\\
&-&           \bigg( -4\,u + 
             {\frac{u\,\left( 2 + u\,{v^2} \right) }{1 - u + u\,v}} - 
             {\frac{1 + {u^2}}
               {\left( v-a \right) \,\left( 1 - u + u\,v \right) \,
                 \left( 1 - {x_{B}} \right) }} + 
             {\frac{\left( 1 - u \right) \,{u^2}\,\left( 1 - v \right) }
               {{{\left( u - {x_{B}} \right) }^2}}} - 
             {\frac{\left( 1 - 2\,u \right) \,u}{u - {x_{B}}}}\nonumber\\ 
&+&              {\frac{2\,{u^2}\,\left( 1 - v \right) }
               {u - {x_{B}}}}
+             {\frac{1 + u\,v}
               {1 - u + u\,v}}\,{p_{g\leftarrow q}}(v) \bigg) \,
\left( \log\left({\frac{\left( 1 - u \right) \,\left( 1 - v \right) }
                {\left( 1 - a \right) \,u}}\right) + 
\log\left({\frac{v-a}{1 - a}}\right)\right)  \nonumber\\
&+&{\frac{1 }{1 - u}}\,\left( \log\left({\frac{v-a}
                    {\left( 1 - a \right) \,v}}\right) - 
                \log\left({\frac{u - {x_{B}}}{1 - {x_{B}}}}\right)
                 \right)\,{p_{g\leftarrow q}}(v)
  \Bigg]  \Bigg\}\,,\\ 
\nonumber\\
D_{2gq,M}^{(1)}&=&
{T_F}\,\Bigg\{
     \Bigg[ \left( 1 - u \right) \,u\,
          \left(\log\left({\frac{1 - u}{\left( 1 - a \right) \,u}}\right)
-1 \right) 
             + {\frac{1}{4}}{\left( {\frac{{{\pi }^2}}{3}} + 
             {{\log\left({\frac{1 - u}{\left( 1 - a \right) \,u}}\right)}^2}
 \right) \,{p_{\overline{q}q\leftarrow g}}(u)} \Bigg]\,
\delta (1 - v)\,\delta (w)\nonumber\\ 
&+&    \delta (wr-w)\,\Bigg[ 
       {\frac{1}{2}}\,{\left({{\frac{\log (1 - v)}{1 - v}}}\right)
_{v[0,\underline{1}]}}\,{p_{\overline{q}q\leftarrow g}}(u)
+ \left(  \left(1 - u \right) \,u   + 
          {\frac{1}{2}}\,{p_{\overline{q}q\leftarrow g}}(u) \,
              \log\left({\frac{1 - u}{\left( 1 - a \right) \,u}}\right)
 \right) \,
        {\left({{\frac{1}{1 - v}}}\right)_{v[0,\underline{1}]}}\nonumber\\
&+&{\frac{1}{2\,\left(v-a \right) }}+\left(
          {\frac{1}{2\,\left(v-a \right)}}\,{p_{\overline{q}q\leftarrow g}}(u)
 -{\frac{1}{1 - a}}\right)\,\left(\log\left({\frac{\left( 1 - u \right) \,
                  \left( 1 - v \right) }{\left( 1 - a \right) \,u}}\right) + 
             \log\left({\frac{v-a}{1 - a}}\right)-1 \right)\nonumber\\
&+&{\frac{1}
         {2\,\left( 1 - v \right) }}\,
\log\left({\frac{v-a}{1 - a}}\right)\,{p_{\overline{q}q\leftarrow g}}(u)
\Bigg]  \Bigg\}\,, 
\end{eqnarray}
where
\begin{eqnarray}
p_{gq\leftarrow q}(x)&=&2\,\frac{1}{1-x}-1-x\,,\\
p_{\overline{q}q\leftarrow g}(x)&=&1-2 x+2x^2\,,\\
p_{g\leftarrow q}(x)&=&\frac{2}{x}-2+x\,.
\end{eqnarray}

\section*{Appendix II}

The redefinition of fracture functions can only factorize singularities in 
the forward region. In the hadronic tensor these singularities show up 
after the angular integration as $(1-w)^{-1+\epsilon}$ factors which have 
to be prescribed as explained in section III. As we mentioned there, special 
care has to be taken with overlapping singularities. In this case, 
inspection of figure \ref{fig:muzub} shows that the only problematic 
configurations are terms singular along $w=1$ and $w=w_r$ in regions 
B0 and B1. Using the procedure described in section III we obtained suitable 
prescriptions in both regions:     
\begin{eqnarray}
(1-w)^{-1+\epsilon_{1}}(w_r-w)^{-1+\epsilon_{2}}&\stackrel{\mbox{B0}}
{\longrightarrow}&
	\frac{1}{\epsilon_{1}\,(\epsilon_{1}+\epsilon_{2})}\,
\frac{\Gamma(1+\epsilon_{1})\,\Gamma(1-\epsilon_{1}-\epsilon_{2})}
{\Gamma(1-\epsilon_{2})}\,\delta(1-w)\,
\delta(a-v)(a-z)^{\epsilon_{1}+\epsilon_{2}}
\left(a\,(1-a)\right)^{1-\epsilon_{1}-\epsilon_{2}}
\nonumber\\
&+&\frac{1}{\epsilon_{1}}\,\delta(1-w)\left((a-v)^{-1+\epsilon_{1}+
\epsilon_{2}}
\right)_{+v[z,\underline{a}]}\,\left(v\,(1-a)\right)^{1-\epsilon_{1}-
\epsilon_{2}}\,w_r^{-\epsilon_{1}}\nonumber\\
&\times&{}_{2}F_{1}\left[\epsilon_{1},\epsilon_{1}+\epsilon_{2},1+\epsilon_{1};
\frac{1}{w_r}\right]
+\left((1-w)^{-1+\epsilon_{1}}(w_r-w)^{-1+\epsilon_{2}}\right)_
{+w[0,\underline{1}]}\,\label{eq:prescr1}\\
(1-w)^{-1+\epsilon_{1}}(w_r-w)^{-1+\epsilon_{2}}&\stackrel{\mbox{B1}}
{\longrightarrow}&
	\frac{1}{\epsilon_{2}\,(\epsilon_{1}+\epsilon_{2})}\,
\frac{\Gamma(1+\epsilon_{2})\,\Gamma(1-\epsilon_{1}-\epsilon_{2})}
{\Gamma(1-\epsilon_{1})}\,\delta(1-w)\,
\delta(v-a)(1-a)^{\epsilon_{1}+\epsilon_{2}}
\left(a\,(1-a)\right)^{1-\epsilon_{1}-\epsilon_{2}}
\nonumber\\
&+&\frac{1}{\epsilon_{2}}\,\delta(w_r-w)
\left((v-a)^{-1+\epsilon_{1}+\epsilon_{2}}
\right)_{+v[\underline{a},1]}\,\left(v\,(1-a)\right)^{1-\epsilon_{1}-\epsilon_
{2}}\,w_r^{-\epsilon_{2}}\nonumber\\
&\times&{}_{2}F_{1}\left[\epsilon_{2},\epsilon_{1}+\epsilon_{2},
1+\epsilon_{2};w_r\right]
+\left((1-w)^{-1+\epsilon_{1}}(w_r-w)^{-1+\epsilon_{2}}\right)_
{+w[0,\underline{w_r}]}\label{eq:prescrw2}
\end{eqnarray} 
where $\epsilon_{1}$ and $\epsilon_{2}$ are multiples of the regulator 
$\epsilon$. Notice that these expressions are valid to all orders in 
$\epsilon$.
Terms singular only along $w=1$ can be prescribed using the standard rule in
eq. (\ref{eq:prescu1}). Terms singular in $w=1$ and $w=0$ can be managed
by partial fractioning:
\begin{equation}
\frac{1}{1-w}\,\frac{1}{w}=\frac{1}{1-w}+\frac{1}{w}
\end{equation} 
and be prescribed also as in eq. (\ref{eq:prescu1}). 

\section*{Appendix III}
The singular pieces of the order $\alpha_s^2$ partonic cross sections 
$\hat{\sigma}_{gg}$ and $\hat{\sigma}_{gq}$ in region B0 can be 
written as
\begin{eqnarray}\label{eq:sigma2gq}
\left.d\hat{\sigma}^{(2)}_{gg,M}\right|_{\mbox{\tiny B0}}
	&=&\sum_{q}c_q\,C_{\epsilon}^{\,2}
	\Biggl\{
	\frac{1}{\epsilon^2}\,\biggl[{}8\,P^{(0)}_{q\leftarrow g}(u)\,
	P^{(0)}_{g\leftarrow q}(v)\,\delta(w)+
	4\Bigl(P^{(0)}_{q\leftarrow g}(u)
	\otimes \tilde{P}^{(0)}_{gq\leftarrow q}(u,v)
	+P^{(0)}_{g\leftarrow q}(v)
	\otimes \tilde{P}^{(0)}_{\bar{q}q\leftarrow g}(u,v)\nonumber\\
&+&	P^{(0)}_{q\leftarrow g}(u)
	\otimes^{\prime} \tilde{P}^{(0)}_{gg\leftarrow g}(u,v)
	\Bigr)\delta(1-w)
	\biggr]
	+\frac{1}{\epsilon}\,\biggl[2\,P^{(1)}_{gq\leftarrow g}(u,v)
	\,\delta(1-w)
	+2\,P^{(0)}_{q\leftarrow g}(u)\otimes C^{(1)}_{1qg,M}(u,v,w)
	\nonumber\\
&+&2	\,P^{(0)}_{g\leftarrow q}(v)\otimes C^{(1)}_{1gq,M}(u,v,w)
	+2\,\frac{1-x_B}{x_B}\,\tilde{P}^{(0)}_{gg\leftarrow g}(u,v)
	\otimes^{\prime}C^{(1)}_{g,M}(u)\,\delta(1-w)
	\biggr]
	+{\cal O}(\epsilon^{0})
	\Biggr\}\nonumber\\
\nonumber\\
\left.d\hat{\sigma}^{(2)}_{gq,M}\right|_{\mbox{\tiny B0}}
	&=&c_q\,C_{\epsilon}^{\,2}
	\Biggl\{
	\frac{1}{\epsilon^2}\,\biggl[
	2\Bigl(
	\tilde{P}^{(0)}_{\bar{q}q\leftarrow g}(u,v)\otimes^{\prime}
	P^{(0)}_{q\leftarrow q}(u)+
	\tilde{P}^{(0)}_{\bar{q}q\leftarrow g}(u,v)\otimes
	P^{(0)}_{q\leftarrow q}(v)
	+\tilde{P}^{(0)}_{\bar{q}q\leftarrow g}(u,v)\otimes
	P^{(0)}_{g\leftarrow g}(u)\nonumber\\
&-&	\frac{1}{2}\beta_{0}\,\tilde{P}^{(0)}_{\bar{q}q\leftarrow g}(u,v)
	\Bigr)\delta(1-w)
	+4\,P^{(0)}_{q\leftarrow g}(u)
	\, P^{(0)}_{q\leftarrow q}(v)\,\delta(w)\biggr]
	+\frac{1}{\epsilon}\biggl[P^{(1)}_{\bar{q}q\leftarrow g}(u,v)\,
	\delta(1-w)\nonumber\\
&-&	\beta_{0}\,C^{(1)}_{1gq,M}(u,v,w)
	+2\,P^{(0)}_{g\leftarrow g}(u)\otimes C^{(1)}_{1gq,M}(u,v,w)
	+2\,P^{(0)}_{q\leftarrow q}(v)\otimes C^{(1)}_{1gq,M}(u,v,w)
	\nonumber\\
	&+&2\,\frac{1-x_B}{x_B}\,\tilde{P}^{(0)}_{\bar{q}q\leftarrow g}(u,v)
	\otimes^{\prime} C^{(1)}_{q,M}(u)\,\delta(1-w)
	\biggr]
	+{\cal O}(\epsilon^{0})
	\Biggr\}\\
\end{eqnarray} 
where the $\tilde{P}^{(0)}_{ki\leftarrow j}(u,v)$ are defined in 
eq.(\ref{eq:kertilde}) and the functions $C^{(1)}_{i,M}(u)$ are the 
coefficient functions of totally inclusive DIS at ${\cal O}(\alpha_s)$, 
they can be found in refs. \cite{grau,zijli}. 
Convolutions between kernels are as in 
eq. (\ref{eq:conv}) with the replacement $v\rightarrow(1-x_B)\,v/x_B$ whereas 
the convolutions between kernels and coefficient functions are given by
\begin{eqnarray}
P_{i\leftarrow j}(u)\otimes C(u,v,w)&=&\int_{\frac{x_B}{x_B+(1-x_B)\,v}}^
{\frac{x_B}{x_B+(1-x_B)\,z}}\frac{d\bar{u}}{\bar{u}}\,P_{i\leftarrow j}
\left(\frac{u}{\bar{u}}\right)\,C(\bar{u},v,w)\,,\nonumber\\
P_{i\leftarrow j}(v)\otimes C(u,v,w)&=&\int_{a(u)}^{1}\frac{d\bar{v}}{v}\,
P_{i\leftarrow j}\left(\frac{v}{\bar{v}}\right)\,C(u,\bar{v},w)\,,\\
\tilde{P}_{ki\leftarrow j}(u,v)\otimes^{\prime}C(u)&=&
\int_{\frac{x_B-u\,v\,(1-x_B)}{x_B}}^{1}\,\frac{u}{\bar{u}}\,
\frac{d\bar{u}}{\bar{u}}\tilde{P}_{ki\leftarrow j}\left(\bar{u},
\frac{v\,u\,(1-x_B)}{x_B\,\bar{u}}\right)C\left(\frac{u}{\bar{u}},v,w\right)
\end{eqnarray}


\end{fmffile}

\begin{thebibliography}{3}
\bibitem{ven}L. Trentadue and G. Veneziano, {\em Phys. Lett.}
{B323}, (1993) 201.
\bibitem{grau} D. Graudenz, {\em Nuc. Phys.} B432 351 (1994)
\bibitem{massi} M. Grazzini, L. Trentadue, G. Veneziano, {\em Nucl. Phys.} B519, (1998) 394;\\
 M. Grazzini, {\em Phys. Rev.} {D57}, (1998) 4352.
\bibitem{npb} D. de Florian, C. A. Garc\'{\i}a Canal and R. Sassot, {\em
Nuc. Phys.} B470,  (1996) 195.
\bibitem{ppp} see C.A. Garcia Canal, R. Sassot, {\em Int. Jour. Mod. Phys.} 
A15  (2000) 3587 and references therein.
\bibitem{prd98} D. de Florian and R. Sassot, {\em Phys. Rev.} {D58}, (1998) 054003.
\bibitem{dss} D. de Florian, O. A. Sampayo, R. Sassot, {\em Phys. Rev.} {D57}
 (1998) 5803.   
\bibitem{ds}D. de Florian, R. Sassot, {\em Phys. Rev.} {D62}  (2000) 094025.
\bibitem{prd97} D. de Florian and R. Sassot, {\em Phys. Rev.} {D56}, (1997) 426.
\bibitem{zijli} E.B. Zijlstra, W.L. van Neerven, {\em Nuc.Phys.} B383  (1992) 525.
\bibitem{solo} W.L. van Neerven, {\em Nuc.Phys.} B268  (1986) 453.
\bibitem{been} W. Beenakker, H. Kuijf, W. L. van Neerven and J. Smith, {\em Phys. Rev.} D40 (1989) 54. W. Beenakker, Ph. D. Thesis, Leiden, 1989. 
\bibitem{ingo} See for example,  {\em NLO QCD Corrections to Polarized Photo
and
Hadroproduction of Heavy Quarks}, I. Bojak Ph.D Thesis, Universitet Dortmund
(2000).
\bibitem{glov} E. W. N. Glover, hep-ph/0211412.
\bibitem{Bol} C. G. Bollini, J. J. Giambiagi, {\em Nuovo Cimento} 12B,
(1972) 20, G. 't Hooft, M. Veltman,  {\em Nucl. Phys.} B44, (1972) 189.
\bibitem{AP} V. N. Gribov, L. N. Lipatov, {\em Sov. J. Nucl. Phys.} 15
(1972) 438, G. Altarelli, G. Parisi, {\em Nucl. Phys.} B126 (1977) 298.
\bibitem{math} S. Wolfram, {\em Mathematica} Third Edition, 
Cambridge University Press 1996.  
\bibitem{tracer} M. Jamin, M.E. Lautenbacher, {\em Comp. Phys. Comm.} 74 
(1993) 265.
\bibitem{pasa} G. Passarino, M. Veltamn, {\em Nucl. Phys.} B160 (1979) 151.
\bibitem{rijken} P. J. Rijken, W. L  van Neerven, {\em Phys. Lett.} B386 
(1992) 422.
\bibitem{arnold} P. B. Arnold, R. P. Kauffman, {\em Nucl. Phys.} B349 (1991) 
381.
\bibitem{hamberg} R. Hamberg, W. L. van Neerven, T. Matsuura, {\em Nucl. Phys.}
 B359 (1991) 343. T. Matsuura, Ph. D. Thesis, Leiden, 1989.
\bibitem{GRV} M. Gl\"uck, E. Reya, A. Vogt, {\em Eur. Phys. J.} C5 (1998) 461. 
\bibitem{kretzer} S. Kretzer, {\em Phys. Rev.} D62 (2000) 054001.
\end{thebibliography}
\end{document}